\documentclass[sigconf]{acmart}

\AtBeginDocument{%
  \providecommand\BibTeX{{%
    \normalfont B\kern-0.5em{\scshape i\kern-0.25em b}\kern-0.8em\TeX}}}

\copyrightyear{2024} 
\acmYear{2024} 
\setcopyright{acmlicensed}\acmConference[KDD '24]{Proceedings of the 30th ACM SIGKDD Conference on Knowledge Discovery and Data Mining}{August 25--29, 2024}{Barcelona, Spain}
\acmBooktitle{Proceedings of the 30th ACM SIGKDD Conference on Knowledge Discovery and Data Mining (KDD '24), August 25--29, 2024, Barcelona, Spain}
\acmDOI{10.1145/3637528.3671820}
\acmISBN{979-8-4007-0490-1/24/08}

\usepackage{amsmath,amsfonts,dsfont,multirow,multicol,epsfig,url,array,makecell,balance,color,epstopdf,amsthm}
\usepackage[normalem]{ulem}
\usepackage{algorithmic}
\usepackage[ruled, vlined, linesnumbered]{algorithm2e}
\usepackage{hhline}
\usepackage{wrapfig}
\usepackage{array}
\usepackage{enumerate}
\usepackage{enumitem}
\usepackage{subfigure}
\usepackage{booktabs}
\usepackage{xcolor,colortbl}
\usepackage{bm}
\usepackage{lipsum}
\usepackage{tabularx}
\usepackage{diagbox}
\usepackage{caption}
\usepackage{threeparttable}
\SetCommentSty{mycommfont}
\usepackage[utf8]{inputenc} 
\usepackage[T1]{fontenc}    
\usepackage{hyperref}       
\usepackage{booktabs}       
\usepackage{nicefrac}       
\usepackage{microtype}      
\usepackage{comment}
\usepackage{tcolorbox}
\usepackage{titlesec}
\usepackage[title]{appendix}

\setlist[itemize]{leftmargin=*}
\setlist[enumerate]{leftmargin=*}

\newtheorem{theorem}{Theorem}

\def\header{\vspace{0.8mm} \noindent}

\def\tblcapup{\vspace{0mm}}

\newcommand{\pushright}[1]{\ifmeasuring@#1\else\omit\hfill$\displaystyle#1$\fi\ignorespaces}
\newcommand{\pushleft}[1]{\ifmeasuring@#1\else\omit$\displaystyle#1$\hfill\fi\ignorespaces}

\def\rela{c}
\def\pf{p_f}

\def\epi{\bm{\hat{\pi}}}

\def\vpi{\bm{\pi}}

\def\vap{p}

\def\vq{q}
\def\indmedian{\gamma}
\def\nm{n_m}

\def\A{\mathbf{A}}
\def\D{\mathbf{D}}

\def\calH{\mathcal{H}}
\def\G{\mathcal{G}}

\def\ind{\chi}

\def\E{\mathrm{E}}
\def\Var{\mathrm{Var}}

\def\dmin{d_{\mathrm{min}}}
\def\dmax{d_{\mathrm{max}}}
\def\N{\mathcal{N}}

\def\deg{\textup{\textsc{deg}}\xspace}
\def\neighbor{\textup{\textsc{neigh}}\xspace}
\def\jump{\textup{\textsc{jump}}\xspace}
\def\randint{\mathrm{randint}}

\def\power{\textup{\texttt{PowerIteration}}\xspace}
\def\bpush{\textup{\texttt{ApproxContributions}}\xspace}
\def\bpushpeter{\textup{\texttt{BackwardPush}}\xspace}
\def\mc{\textup{\texttt{MC}}\xspace}
\def\fastppr{\textup{\texttt{FastPPR}}\xspace}
\def\bippr{\textup{\texttt{BiPPR}}\xspace}
\def\unbippr{\textup{\texttt{Undir-BiPPR}}\xspace}
\def\rbs{\textup{\texttt{RBS}}\xspace}
\def\setpush{\textup{\texttt{SetPush}}\xspace}
\def\backmc{\textup{\texttt{BackMC}}\xspace}
\def\sublinear{\textup{\texttt{BPPPush}}\xspace}
\def\sublinearplus{$\textup{\texttt{BPPPush}}^*$\xspace}
\def\samplenode{\textup{\texttt{SampleNode}}\xspace}

\begin{document}

\title{Revisiting Local PageRank Estimation on Undirected Graphs: Simple and Optimal}
\subtitle{[Technical Report]}

\author{Hanzhi Wang}
\affiliation{%
  \institution{Renmin University of China}
  \city{Beijing}
  \country{China}
}
\email{hanzhi_wang@ruc.edu.cn}



\begin{abstract}
We propose a simple and optimal algorithm, \backmc, for local PageRank estimation in undirected graphs: given an arbitrary target node $t$ in an undirected graph $G$ comprising $n$ nodes and $m$ edges, \backmc accurately estimates the PageRank score of node $t$ while assuring a small relative error and a high success probability. 
The worst-case computational complexity of \backmc is upper bounded by $O\left(\frac{1}{\dmin}\cdot \min\left(d_t, m^{1/2}\right)\right)$, where $\dmin$ denotes the minimum degree of $G$, and $d_t$ denotes the degree of $t$, respectively. 
Compared to the previously best upper bound of $ O\left(\log{n}\cdot \min\left(d_t, m^{1/2}\right)\right)$ (VLDB '23), which is derived from a significantly more complex algorithm and analysis, our \backmc improves the computational complexity for this problem by a factor of $\Theta\left(\frac{\log{n}}{\dmin}\right)$ with a much simpler algorithm. Furthermore, we establish a matching lower bound of $\Omega\left(\frac{1}{\dmin}\cdot \min\left(d_t, m^{1/2}\right)\right)$ for any algorithm that attempts to solve the problem of local PageRank estimation, demonstrating the theoretical optimality of our \backmc. We conduct extensive experiments on various large-scale real-world and synthetic graphs, where \backmc consistently shows superior performance.

\end{abstract}



\maketitle

\section{Introduction} \label{sec:intro}
PageRank is a celebrated metric for assessing node centrality in graphs, originally introduced by Google for ranking web pages based on their prominence within the web network~\cite{page1999pagerank, brin1998anatomy}. Over the past two decades, PageRank has evolved into one of the most popular graph centrality metrics, with widespread applications across diverse fields. These include social network analysis~\cite{gupta2013wtf}, spam detection~\cite{gyongyi2004combating_pagerank_spam}, recommender systems~\cite{gori2007itemrank}, graph representation learning~\cite{chen2020GBP, klicpera2019APPNP}, chemical informatics~\cite{mooney2012molecularnetworks}, and bioinformatics~\cite{morrison2005generank}, and more~\cite{gleich2015beyondtheweb}. 
The computation of PageRank has become a fundamental aspect of modern network analysis.

In recent years, the exponential growth in network sizes has sparked significant interest in the field of {\em local} PageRank estimation~\cite{chen2004local, bar2008reversePageRank, setpush2023VLDB, bressan2018sublinear, bressan2023sublinear, lofgren2014FastPPR, lofgren2016BiPPR, lofgren2015bidirectional_undirected, wang2020RBS, lofgren2013personalized, andersen2007contribution}. This problem focuses on the approximation of a given target node's PageRank score, with the goal of exploring only a small portion of the graph. Various practical applications illustrate the utility of local PageRank estimation. For example, in recommender systems, there is a growing need to quickly approximate the PageRank scores of individual users during online operations, as opposed to performing time-consuming global computations of determining all nodes' PageRank scores within a graph~\cite{chen2004local}. Similarly, in web search scenarios, website owners who are interested in enhancing their search engine rankings may only seek the PageRank score of their specific websites, not those of the entire web~\cite{bar2008reversePageRank}. Additionally, in social networks, users often desire to gauge their PageRank-based popularity by efficiently probing the friendship graph, rather than having to traverse the entire networkr~\cite{setpush2023VLDB, bar2008reversePageRank}. Therefore, it is imperative to have highly efficient algorithms for local PageRank estimation.

This paper focuses on local PageRank estimation in {\em undirected} graphs~\footnote{It's important to note, as formally established in~\cite{grolmusz2015note}, that the PageRank scores in undirected graphs are not simply proportional to node degrees, despite common misconceptions to the contrary often cited in the literature. }. We aim to address the problem of computing, with probability $(1-\pf)$, a multiplicative $(1 \pm \rela)$-approximation of the PageRank score for a given target node $t$ in an undirected graph $G$ comprising $n$ nodes and $m$ edges. 
Both $\rela$ and $\pf$ are constants set within the $(0,1)$ range (e.g., $\rela=\pf=0.1$). 
This problem is of great importance from both theoretical and practical aspects. 

\header{\bf Theoretical Motivations. } Existing studies on local PageRank estimation can be broadly divided into two categories. The first category focuses on the approximation of a single node's PageRank score on general {\em directed} graphs~\cite{bressan2018sublinear, bressan2023sublinear, wang2020RBS, lofgren2014FastPPR, lofgren2016BiPPR}, while the second specifically targets {\em undirected} graphs~\cite{setpush2023VLDB, lofgren2015bidirectional_undirected}. For directed graphs, the best-known {\em lower bound} of $\Omega\left(\min\left(n^{1/2} \dmax^{1/2}, n^{1/3} m^{1/3} \right)\right)$ for the worst-case computational complexity was established by Bressan, Peserico, and Pretto~\cite{bressan2018sublinear, bressan2023sublinear}, where $\dmax$ denotes the maximum outdegree of $G$. This lower bound indicates the improbability of achieving a complexity bound of $O(\sqrt{n})$ even on very sparse directed graphs. 
Surprisingly, on undirected graphs, a recent work~\cite{setpush2023VLDB} shows that the computational complexity of local PageRank approximation can be improved to $O\left(\log{n}\cdot \min\left(d_t, m^{1/2}\right)\right)$ by leveraging the symmetry of PageRank vectors in undirected graphs~\footnote{
For readability, we hide multiplicative factors depending on the parameters $\rela, \pf$ and $\alpha$ in the $O()$ notation, following~\cite{bressan2018sublinear, bressan2023sublinear, setpush2023VLDB, wang2020RBS}.
These factors in our upper bound can be found in Section~\ref{sec:algorithm}. }. 
Here, $d_t$ denotes the degree of the target node $t$ in the undirected graph $G$. We note that this upper bound is even asymptotically better than the lower bound established for directed graphs. This encouraging result underscores the theoretical significance and necessity of exploring the complexity bounds of local PageRank estimation on undirected graphs. However, the question of whether this upper bound can be further improved remains open due to the lack of lower bounds for undirected graphs~\cite{setpush2023VLDB}. To the best of our knowledge, only a trivial lower bound of $\Omega(1)$ has been established over the years. This gap in understanding motivates our exploration in this area.

\header{\bf Applications Significance. } 
Local estimation of PageRank scores on undirected graphs is a versatile tool with a range of applications. A typical example is its use in Graph Neural Networks (GNNs). The majority of GNN models are primarily designed for undirected graphs, aiming to derive low-dimensional latent representations of all training nodes from structural and feature information. A message-passing mechanism is typically employed in existing GNN models for feature propagation. In particular, a line of research~\cite{chen2020GBP, klicpera2019APPNP, Bojchevski2020PPRGo, wang2021AGP} utilizes PageRank queries for efficient feature propagation, 
where the initial feature vector serves as the preference vector for PageRank computations (as detailed in Section~\ref{sec:pre}). 
In these models, feature propagation is essentially akin to computing PageRank scores for a selected group of nodes in semi-supervised learning tasks, like community classification on a billion-scale graph Friendster~\cite{chen2020GBP}, where feature vector entries are binary and the number of training nodes is small. Thus, an efficient local PageRank algorithm can greatly enhance the scalability of GNN models in these contexts. 
Moreover, the simulation of random walks in undirected graphs is a prevalent approach widely adopted in various graph learning tasks~\cite{chen2020GBP, Zeng2020graphsaint,wang2021AGP,perozzi2014deepwalk}. While prior studies have demonstrated the empirical efficacy of this method, they fall short of providing a theoretical foundation for the optimality of generating random walks in undirected settings. This paper establishes that merely generating random walks from a specified target node achieves the optimal complexity for local PageRank computation. Our findings aim to lay a theoretical foundation for the efficacy of random walk simulations in undirected graphs, thereby informing the design of graph learning methods tailored to such graphs.

\subsection{Our Contributions} 
In this paper, we address the problem of locally estimating the PageRank score of a target node $t$ in an undirected graph $G$. We achieve the following contributions:
\begin{itemize}
\item 
We introduce \backmc, an algorithm that achieves the worst-case computational complexity of $O\left(\frac{1}{\dmin} \cdot \min\left(d_t, m^{1/2}\right)\right)$, where $\dmin$ represents the minimum degree of the graph $G$, $d_t$ denotes the degree of the target node $t$, and $m$ is the total number of edges in $G$. This computational complexity notably improves upon the previous best upper bound of $O\left(\log{n} \cdot \min\left(d_t, m^{1/2}\right)\right)$, as documented for the \setpush algorithm~\cite{setpush2023VLDB}, by a factor of $\Theta\left(\frac{\log{n}}{\dmin}\right)$. 
Remarkably, although the complexity result of \setpush is derived from a significantly complex analysis, the algorithm structure and theoretical analysis of our \backmc exhibit a surprising simplicity.

\item 
We improve the lower bound for this problem from a trivial bound of $\Omega(1)$ to $\Omega\left(\frac{1}{\dmin}\cdot \min\left(d_t, m^{1/2}\right)\right)$. The matching upper and lower bounds demonstrate that our \backmc has been optimal. 


\item Beyond its theoretical optimality, \backmc distinguishes itself with a clean algorithm structure and straightforward implementation, and thus achieves exceptional empirical performance. 
We conduct extensive experiments on large-scale real-world and synthetic graphs, and \backmc consistently outperforms all baseline algorithms. Notably, it surpasses \setpush in both efficiency and accuracy by up to three orders of magnitude. 

\end{itemize}

\section{Preliminaries} \label{sec:pre}
We denote the underlying undirected graph as $G=(V,E)$, with $\A$ and $\D$ representing the adjacency and degree matrices of $G$, respectively. The graph consists of $n=|V|$ nodes and $m=|E|$ edges. For any undirected edge $(u,v)$ in $G$, nodes $u$ and $v$ are termed as neighbors. For a node $u$ in $G$, the set of its neighbors is denoted as $\N(u)$, and its degree is represented by $d_u=|\N(u)|$. We further define $\dmax=\max_{u\in V}d_u$ and $\dmin=\min_{u\in V}d_u$ as the maximum and minimum degrees of $G$, respectively. We list frequently used notations in Table~\ref{tbl:def-notation} for quick reference. 


\begin{table} [t]
\centering
\renewcommand{\arraystretch}{1.3}
\begin{small}
\tblcapup
\caption{Table of notations.}\label{tbl:def-notation}
\vspace{-2mm}
\begin{tabular} 
{ll} \toprule
{\bf Notation} &  {\bf Description}  \\ \midrule
$G=(V,E)$ & undirected graph with node set $V$ and edge set $E$ \\ 
$\A, \D$ & adjacency and degree matrices of $G$\\ 
$n, m$ & numbers of nodes and edges in $G$ \\ 
$\N(t)$ & set of neighbors of $t$\\ 
$d_t$ & degree of node $t$ \\ 
$\dmax$ & maximum degree of $G$\\ 
$\dmin$ & minimum degree of $G$\\
$\vpi$   & PageRank vector\\ 
$\vpi(t),\epi(t)$   & real and estimated PageRank scores of node $t$ \\ 
$\vpi(s,t)$    & Personalized PageRank score of node $t$ w.r.t node $s$ \\ 
$\alpha$	& teleport probability in defining PageRank \\ 
$\rela$ & relative error parameter\\
$\pf$ & failure probability parameter\\
\bottomrule
\end{tabular}
\end{small}
\end{table}

\subsection{PageRank}
The PageRank vector $\vpi$ of graph $G$ is defined as the stationary distribution of the PageRank Markov chain~\cite{page1999pagerank, brin1998anatomy}, satisfying: 
\begin{align}~\label{eqn:def_pagerank}
\vpi=(1-\alpha)\A\D^{-1}\cdot \vpi +\frac{\alpha}{n}\cdot \bm{1}, 
\end{align}
where $\alpha \in (0,1)$ is the teleport probability, and $\bm{1}$ is an all-one column vector of length $n$. The vector $\bm{1}/n$ is termed as the preference or personalized vector in PageRank. 

For any node $u$ in $G$, the PageRank score of $u$, denoted by $\vpi(u)$, corresponds to the entry in $\vpi$ associated with node $u$. 
It is well known that $\vpi(u)$ equals the probability that an $\alpha$-discounted random walk, starting from a random source node uniformly chosen in $V$, terminates at node $u$~\cite{page1999pagerank, brin1998anatomy}. Here, an $\alpha$-discounted random walk is a type of random walk where the length of the walk is a random variable that takes on value $\ell$ with probability $\alpha (1-\alpha)^\ell$ for each $\ell \ge 0$. In other words, at each step of the $\alpha$-discounted random walk, there is a probability of $\alpha$ that the walk will terminate, and a probability of $(1-\alpha)$ that the walk will continue to the next step. 

Equation~\eqref{eqn:def_pagerank} implies that the following recursive equality holds for any node $u\in V$.  
\begin{align}\label{eqn:ite_pagerank}
\vpi(u)=\sum_{v\in \N(u)}\frac{(1-\alpha)\vpi(v)}{d_v}+\frac{\alpha}{n}. 
\end{align}
Moreover, Wang and Wei~\cite{setpush2023VLDB} demonstrate that for any node $u\in V$, 
\begin{align}\label{eqn:pagerank_lowerbound_max}
\vpi(u)\ge \max\left(\frac{\alpha}{n}, \frac{\alpha d_u \sqrt{2(1-\alpha)} }{n \sqrt{m}}\right). 
\end{align}
We present the proof below for the sake of completeness. 
\begin{proof}[Proof of Equation~\eqref{eqn:pagerank_lowerbound_max}]
By Equation~\eqref{eqn:ite_pagerank}, we have $\vpi(u)\ge \alpha /n$ for any node $u\in V$. Applying $\vpi(u)\ge \alpha /n$ into Equation~\eqref{eqn:ite_pagerank} further yields that
\begin{align}\label{eqn:mid_lower}
\vpi(u)\ge \sum_{v\in \N(u)}\frac{(1-\alpha)}{d_v}\cdot \frac{\alpha}{n}+\frac{\alpha}{n}=\frac{\alpha}{n}\cdot \left(\sum_{v\in \N(u)}\frac{(1-\alpha)}{d_v}+1\right). 
\end{align}
In particular, by the Cauchy-Schwarz inequality, we have
\begin{align*}
\left(\sum_{v\in \N(u)}\frac{1}{d_v}\right)\cdot \left(\sum_{v\in \N(u)}d_v\right) \ge \left(\sum_{v\in \N(u)}1\right)^2=d_u^2. 
\end{align*}
It is important to note that $\sum_{v\in \N(u)}d_v \le 2m$, therefore giving that $\sum_{v\in \N(u)}1/d_v \ge d_u^2/2m$. By Plugging into Inequality~\eqref{eqn:mid_lower}, we have
\begin{align*}
\vpi(u)\ge \frac{\alpha}{n}\cdot \left(\frac{(1-\alpha)d_u^2}{2m}+1\right)=\frac{\alpha d_u}{n}\cdot \left(\frac{(1-\alpha)d_u+2m/d_u}{2m}\right). 
\end{align*}
By the AM-GM inequality, we further have
\begin{align*}
\vpi(u)\ge \frac{\alpha d_u}{n}\cdot \frac{2\sqrt{2(1-\alpha)m}}{2m}=\frac{\alpha d_u}{n}\cdot \frac{\sqrt{2(1-\alpha)}}{\sqrt{m}}.  
\end{align*}
Combining these results gives the claimed inequality. 
\end{proof}

\subsection{Personalized PageRank}
Personalized PageRank (PPR) serves as an ego-centric counterpart to PageRank, quantifying the probability that an $\alpha$-discounted random walk, starting from a source node $u$, terminates at a target node $v$. This is expressed as the PPR score of $v$ with respect to $u$, symbolized by $\vpi(u,v)$. It follows naturally that the PageRank score of $v$ is the average of $\vpi(u,v)$ across all nodes $u$ in the graph, formalized as 
\begin{align}\label{eqn:PageRank_PPR}
\vpi(v)=\frac{1}{n}\cdot\sum_{u\in V}\vpi(u,v). 
\end{align}

Specifically, on undirected graphs, PPR scores exhibit symmetry for any pair of nodes $(u,v)\in V^2$:  
\begin{align}\label{eqn:birectional_ppr}
\vpi(u,v)\cdot d_u=\vpi(v,u) \cdot d_v. 
\end{align}
The proof of Equation~\eqref{eqn:birectional_ppr} is detailed in~\cite{lofgren2015bidirectional_undirected}.

\subsection{Computational Model}\label{subsec:graphmodel}
In this paper, we adopt the standard RAM model for computational complexity. To establish lower bounds, we consider query complexity under the standard {\em arc-centric graph access model}~\cite{goldreich1998property, goldreich2002property}, where local algorithms can access the underlying graph only through a graph oracle via several local queries and a global $\jump()$ operation. This setting aligns well with practical situations with massive-scale network structures. Specifically, for undirected graphs, the graph oracle supports three elementary query operations, each taking unit time: $\deg(u)$, which returns $d_u$; $\neighbor(u,i)$, which returns the $i$-th node in $\N(u)$; $\jump()$, which returns a random node uniformly chosen from $V$.
Algorithm~\ref{alg:samplenode} is an example on how to leverage these query operations to sample random walks in the graph. The query complexity of a graph algorithm is defined as the number of elementary query operations invoked on the graph oracle. It is worth noting that the query complexity of an algorithm serves as a lower bound for its computational complexity.


\section{Related Work} \label{sec:related}
The problem of estimating PageRank scores locally was introduced in ~\cite{chen2004local}, 
and, in its various forms, has received considerable attention over the past decade~\cite{chen2004local, bar2008reversePageRank, setpush2023VLDB, bressan2018sublinear, bressan2023sublinear, lofgren2014FastPPR, lofgren2016BiPPR, lofgren2015bidirectional_undirected, wang2020RBS, lofgren2013personalized, andersen2007contribution, fogaras2005MC}. 
These methods can be broadly categorized into three groups based on their underlying techniques. 
A summary of their complexity bounds is provided in Table~\ref{tbl:comparison}. In this section, we will briefly review these methods. Section~\ref{sec:algorithm} will delve into the limitations of existing methods and offer a detailed comparison with our \backmc algorithm.

\begin{table} [t]
\centering
\renewcommand{\arraystretch}{1.4}
\begin{small}
\tblcapup
\caption{
Summary of computational complexities for estimating the PageRank score of the target node $\boldsymbol{t}$ in undirected graph $\boldsymbol{G}$. Results only applicable to undirected graphs are marked with $\star$. 
}\label{tbl:comparison}
\vspace{-2mm}
\begin{tabular}{llc} 
\toprule
{\bf Method} &  {\bf Complexity Bound}  & {\bf Notes} \\ \midrule
\power~\cite{page1999pagerank} & $O(m \log{n})$ & \\  
\bpushpeter~\cite{lofgren2013personalized} & $O\left(\log{n}\cdot \min\left(n d_t, m\right)\right)$ & $\star$\\ 
\mc~\cite{fogaras2005MC} & $O(n)$  & \\ 
\rbs~\cite{wang2020RBS} & $O(n\log{n})$  & \\ 
\unbippr~\cite{lofgren2015bidirectional_undirected} & $O\left(n^{1/2} d_t^{1/2}\right)$  & $\star$\\ 
\sublinear~\cite{bressan2018sublinear} & {$O\left(\log{n}\cdot \min\left(n^{3/4} \dmax^{1/4}, n^{5/7} m^{1/7} \right)\right)$} & \\ 
\sublinearplus~\cite{bressan2023sublinear} & {$O\left(\log{n}\cdot \min\left(n^{2/3} \dmax^{1/3}, n^{2/3} m^{1/6} \right)\right)$}  & \\ 
\setpush~\cite{setpush2023VLDB} & {$O\left(\log{n}\cdot \min\left(d_t, m^{1/2}\right)\right)$}  & $\star$ \\ 
\rowcolor{gray!20} \backmc (Ours) & $O\left(\frac{1}{\dmin}\cdot \min\left(d_t, m^{1/2}\right)\right)$ & $\star$ \\ \midrule
\rowcolor{gray!20} Lower Bound (Ours) & $\Omega\left(\frac{1}{\dmin}\cdot \min\left(d_t, m^{1/2}\right)\right)$ & $\star$ \\
\bottomrule
\end{tabular}
\vspace{-2mm}
\end{small}
\end{table}

The first category~\cite{fogaras2005MC, avrachenkov2007monte, borgs2012sublinear, borgs2014multiscale} is inspired by the probabilistic interpretation of PageRank scores. The seminal work by Fogaras et al.~\cite{fogaras2005MC} introduced a Monte Carlo (MC) method that initiates a series of $\alpha$-discounted random walks across the graph $G$, where each random walk is generated from a uniformly random source node in $G$. The MC method calculates the proportion of walks that terminate at the given target node $t$, utilizing this ratio to approximate $\vpi(t)$. To achieve a multiplicative $(1\pm \rela)$-approximation of $\vpi(t)$ with probability $1-\pf$, the expected computational complexity of the MC method is upper bounded by $O(n)$. This result is applicable to both directed and undirected graphs. 

Another line of research~\cite{andersen2007contribution, lofgren2013personalized, wang2020RBS, setpush2023VLDB} focuses on the estimation of $\vpi(s,t)$ for all $s\in V$, using Equation~\eqref{eqn:PageRank_PPR} to derive an estimate for $\vpi(t)$. These algorithms start with one unit of probability mass at node $t$ and perform a sequence of {\em push} operations to distribute the probability mass across the graph $G$. The push operation, applied to node $u$, redistributes the probability mass from node $u$ to its neighbors, following the recurrence equation in Equation~\eqref{eqn:ite_pagerank}. 
The seminal paper~\cite{andersen2007contribution} in this category introduces the \bpush algorithm, which achieves a worst-case computational complexity of $O\left(n d_t\right)$ for determinsitically computing a multiplicative $(1 \pm \rela)$-approximation of $\vpi(t)$. 
It is worth noting that this result is only applicable to undirected graphs. 
On directed graphs, the running time of \bpush can only be bounded by $O\left(m\right)$ in expectation over a uniform random choice of $t\in V$. 
Subsequent studies, such as the \bpushpeter algorithm by Lofgren and Goel~\cite{lofgren2013personalized}, improved the framework of \bpush to achieve a worst-case computational complexity of $O\left(\log{n}\cdot \min\left(n d_t, m\right)\right)$ on undirected graphs and an average running time of $O\left(m\log{n}\right)$ over all $t\in V$ on directed graphs.
A recent advancement, \rbs~\cite{wang2020RBS}, further improves the worst-case computational complexity of this problem to $O\left(n \log{n}\right)$. 
This result applies to both directed and undirected graphs. The key idea of \rbs is to pre-sort the neighbors of each node in ascending order of their degrees and, in each push operation, propagate probability mass only to neighbors with small degrees. 
The \setpush algorithm~\cite{setpush2023VLDB} achieves the best-known worst-case complexity of $O\left(\log{n} \cdot \min\left(d_t, m^{1/2}\right)\right)$ for estimating $\vpi(t)$ on undirected graphs. 
The algorithm critically depends on the symmetry of PPR scores in undirected graphs. It employs a complex push strategy to estimate $\vpi(t,s)$ for all $s\in V$, and then derives the approximation of $\vpi(t)$ using the equation $\vpi(t)=\frac{1}{n}\cdot \sum_{s\in V} \frac{d_t}{d_s}\cdot \vpi(t,s)$, as inferred from Equation~\eqref{eqn:PageRank_PPR} and~\eqref{eqn:birectional_ppr}. 

Additionally, a set of papers presents novel results by combining the Monte Carlo method and push operations. This idea was introduced in the \fastppr algorithm~\cite{lofgren2014FastPPR}, and further refined in the \bippr algorithm~\cite{lofgren2016BiPPR}, both designed for directed graphs. 
An alternative version of \bippr, referred to as \unbippr, has been tailored for undirected graphs. \unbippr achieves a worst-case computational complexity of $O\left(n^{1/2} d_t^{1/2}\right)$ for estimating $\vpi(t)$. 
In the recent work by Bressan, Peserico, and Pretto~\cite{bressan2018sublinear}, a worst-case computational complexity of $O\left(\log{n}\cdot \min\left(n^{3/4} \dmax^{1/4}, n^{5/7} m^{1/7} \right)\right)$ was achieved for general directed graphs. Subsequently, this bound was further improved to $O\left(\log{n}\cdot \min\left(n^{2/3} \dmax^{1/3}, n^{2/3} m^{1/6} \right)\right)$ in~\cite{bressan2023sublinear}, establishing itself as the best-known worst-case complexity bound for local PageRank approximation in directed graphs. 

As for lower bounds, Bar-Yossef and Mashiach~\cite{bar2008reversePageRank} have proven that, in directed graphs, the lack of $\jump()$ operation imposes a lower bound of $\Omega(n)$ on the query complexity for locally approximating $\vpi(t)$. Moreover, Bressan, Peserico, and Pretto~\cite{bressan2018sublinear} have shown that a lower bound of $\Omega\left(\min\left(n^{1/2} \dmax^{1/2}, n^{1/3} m^{1/3} \right)\right)$ elementary query operations is necessary in the worst case to estimate $\vpi(t)$ within a multiplicative $O(1)$ factor with probability $\Omega(1)$ in directed graphs. However, the lower bounds pertaining to the estimation of $\vpi(t)$ on undirected graphs remain unclear.


\section{Algorithm}\label{sec:algorithm}
This section presents our \backmc algorithm. We demonstrate that \backmc can achieve the following upper bound simply by generating a series of $\alpha$-discounted random walks from node $t$. 
\begin{theorem}\label{thm:nranalysis}
Given an undirected graph $G$ and a target node $t\in V$, the expected computational complexity of \backmc for computing a multiplicative $(1\pm \rela)$-approximation of $\vpi(t)$ with probability at least $1-\pf$ is $O\left(\frac{1}{\dmin}\cdot \min\left(d_t, m^{1/2}\right)\right)$. 
\end{theorem}
Algorithm~\ref{alg:MC} provides the pseudocode for \backmc. At the heart of Algorithm~\ref{alg:MC} is the $\samplenode(G, t, \alpha)$ function, responsible for simulating an $\alpha$-discounted random walk from node $t$ in $G$. We define the indicator variable $\ind_v$, which equals $1$ if $\samplenode(G, t, \alpha)=v$. We also compute a temporary estimator 
$\vq(t)=\sum_{v\in V} \frac{d_t}{n\cdot d_v} \cdot \ind_v$. 
We invoke $\samplenode(G, t, \alpha)$ for $n_r$ times to obtain $n_r$ independent realizations of $q(t)$, and take their average as the final estimate $\epi(t)$ for $\vpi(t)$. 







\begin{algorithm}[t]
\DontPrintSemicolon
\caption{$\backmc(G, t, \alpha, n_r)$}
\label{alg:MC}
\BlankLine
\KwIn{undirected graph $G=(V,E)$, target node $t\in V$, damping factor $\alpha$, the number of random walks $n_r$\\}
\KwOut{$\epi(t)$ as an estimate of $\vpi(t)$\\}
$\epi(t)\gets 0$\;
\For{$w$ from $1$ to $n_r$}{
    $v \gets \samplenode(G,t,\alpha)$\; 
    $\epi(t) \gets \epi(t)+\frac{1}{n_r}\cdot \frac{d_t}{n d_v}$ \textcolor{gray}{//$\vq(t)=\frac{d_t}{n d_v}$} \;
}
\Return $\epi(t)$;
\end{algorithm}

\begin{algorithm}[t]
\DontPrintSemicolon
\caption{$\samplenode(G, u, \alpha)$}
\label{alg:samplenode}
\BlankLine
\KwIn{undirected graph $G=(V,E)$, source node $u\in V$, damping factor $\alpha$\\}
\KwOut{a sampled node $v$\\}
$v \gets u$\; 
\While{\textup{true}}{
    with probability $\alpha$ \Return $v$\; 
    $v\gets \neighbor(v,\randint(\deg(v)))$ \textcolor{gray}{//$\randint(\deg(v))$ returns a uniformly random integer in $\left[1,\deg(v)\right]$} \;
}
\end{algorithm}

\subsection{Analysis}
This section presents the proof of Theorem~\ref{thm:nranalysis}. It is straightforward that $\ind_v$ is a Bernoulli random variable that takes on value $1$ with probability $\vpi(t,v)$. Therefore, $\vq(t)$ is an unbiased estimator for $\vpi(t)$ since 
$\vpi(t)=(1/n)\cdot \sum_{v\in V} \vpi(t,v)\cdot (d_t/d_v)$, 
as inferred from Equation~\eqref{eqn:PageRank_PPR} and~\eqref{eqn:birectional_ppr}. It is important to note that \backmc computes $\epi(t)$ by averaging $n_r$ independent realizations of $\vq(t)$. Thus, $\epi(t)$ also emerges as an unbiased estimator for $\vpi(t)$. 

We now proceed to bound the variance of the estimator $\epi(t)$ generated by \backmc. It is well-known that the variance of the Bernoulli random variable $\ind_v$ is given by $\Var\left[\ind_v\right]=\vpi(t,v)(1-\vpi(t,v))\le \vpi(t,v)$.  It is worth noting that for all $v\in V$, the random variables $\ind_v$ are negatively correlated since an $\alpha$-discounted random walk can only terminate at a single node. Consequently, we have
\begin{align*}
\Var\left[\vq(t)\right]\le \sum_{v\in V}\Var\left[\frac{d_t}{n d_v}\cdot \ind_v\right]\le \sum_{v\in V}\frac{d_t^2}{n^2 d_v^2}\cdot \vpi(t,v). 
\end{align*}
By utilizing Equation~\eqref{eqn:birectional_ppr}, we can further derive that
\begin{align*}
\Var\left[\vq(t)\right]\le \sum_{v\in V}\frac{d_t}{n^2 d_v}\cdot \vpi(v,t)\le \frac{d_t}{n^2 \dmin}\cdot \sum_{v\in V}\vpi(v,t)=\frac{d_t \cdot \vpi(t)}{n \cdot \dmin}, 
\end{align*}
where we employ Equation~\eqref{eqn:PageRank_PPR} to assert that $\frac{1}{n}\cdot \sum_{v\in V}\vpi(v,t)=\vpi(t)$. Recall that \backmc computes $\epi(t)$ as the average of $n_r$ independent realizations of $\vq(t)$. As a result, we can bound the variance of the estimator $\epi(t)$ in \backmc as $\Var[\epi(t)]\le \frac{d_t \cdot \vpi(t)}{n \cdot n_r \cdot \dmin}$. 

Therefore, with $n_r=\frac{3 d_t}{\rela^2 n \vpi(t) \dmin}$, Chebyshev’s inequality guarantees that
\begin{align*}
\Pr\left\{ \left|\epi(t)-\vpi(t)\right| \ge \rela \vpi(t)\right\} \le \frac{\Var\left[\epi(t)\right]}{\rela^2 \left(\vpi(t)\right)^2}\le \frac{d_t}{\rela^2 n \vpi(t) \cdot n_r \dmin}=\frac{1}{3}. 
\end{align*}
Given that $\vpi(t)\ge \max\left(\frac{\alpha}{n}, \frac{\alpha d_t \sqrt{2(1-\alpha)} }{n \sqrt{m}}\right)$ (from Inequality~\eqref{eqn:pagerank_lowerbound_max}), we can ensure that $\Pr\left\{ \left|\epi(t)-\vpi(t)\right| \ge \rela \vpi(t)\right\} \le 1/3$ by setting $n_r=\frac{3}{\rela^2 \alpha \dmin}\cdot \min\left(d_t, \frac{\sqrt{m}}{\sqrt{2(1-\alpha)}}\right)$. 

To further reduce the failure probability to $\pf$, we use the Median trick. The core idea is to independently run \backmc multiple times and take the median of the resulting $\epi(t)$ as the final estimate for $\vpi(t)$. Specifically, we use $\nm$ to denote the number of \backmc runs. For each integer $j\in [1,\nm]$, we define the indicator variable $\indmedian_j$, which equals $1$ if the $j$-th run fails to achieve $\left|\epi(t)-\vpi(t)\right| < \rela \cdot \vpi(t)$. Based on the earlier proof, we have $\E\left[\indmedian_j\right]=\Pr\left\{\indmedian_j=1\right\}\le 1/3$ when $n_r=\frac{3}{\rela^2 \alpha \dmin}\cdot \min\left(d_t, \frac{\sqrt{m}}{\sqrt{2(1-\alpha)}}\right)$. Furthermore, we define $\indmedian=\sum_{j=1}^{\nm}\indmedian_j$ as the count of times \backmc fails to produce a satisfactory approximation of $\vpi(t)$. The probability that the median of these estimates is not a multiplicative $(1\pm \rela)$-approximation of $\vpi$ can thus be quantified as $\Pr\left\{\indmedian \ge \nm/2\right\}$. Given that $\E\left[ \indmedian\right]=\sum_{j=1}^{\nm}\E\left[\indmedian_j\right]\le \nm/3$, by Hoeffding's inequality, $\nm=\left\lceil 18\cdot \ln(1/\pf)\right\rceil$ yields:
\begin{align*}
\Pr\left\{\indmedian \ge \frac{\nm}{2}\right\}=\Pr\left\{\indmedian - \E\left[\indmedian\right]\ge\frac{\nm}{6}\right\}\le \exp\left(\frac{-\nm}{18}\right)\le \pf. 
\end{align*}

As a consequence, the total number of $\alpha$-discounted random walks generated in $G$ turns out to be 
\begin{align*}
n_r \cdot \left\lceil 18\cdot \ln(1/\pf)\right\rceil=\frac{54 \cdot \left\lceil\log{(1/\pf)}\right\rceil}{\rela^2 \alpha \dmin}\cdot \min\left(d_t, \frac{\sqrt{m}}{\sqrt{2(1-\alpha)}}\right). 
\end{align*} 
We observe that the expected length of each $\alpha$-discounted random walk is $1/\alpha=O(1)$. Consequently, the expected running time of the entire process can be upper bounded by 
\begin{align*}
\frac{n_r}{\alpha}
&=\frac{54 \cdot \left\lceil\log{(1/\pf)}\right\rceil}{\rela^2 \alpha^2 \dmin}\cdot \min\left(d_t, \frac{\sqrt{m}}{\sqrt{2(1-\alpha)}}\right)\\
&=O\left(\frac{1}{\dmin}\cdot \min\left(d_t, \sqrt{m}\right)\right). 
\end{align*}
This completes the proof of Theorem~\ref{thm:nranalysis}.

\header{\bf Remark. } We note that the setting of $n_r$ requires explicit knowledge of $m$ and $\dmin$. As demonstrated in~\cite{bressan2018sublinear, bressan2023sublinear, dagum2000optimal}, this requirement can be easily waived. To achieve this, we run \backmc with an initial value for $n_r$ (e.g., $n_r=2$). If we find that $n_r \cdot \epi(t)=\Omega(\rela^{-1}\ln{(1/\pf)})$ during the process, standard concentration bounds allow us to safely terminate \backmc and return $\epi(t)$ as a multiplicative $(1\pm \rela)$-approximation of $\vpi(t)$. In case this condition is not met, we double the value of $n_r$ and repeat the process. This modification does not affect the validity of Theorem~\ref{thm:nranalysis}.

\subsection{Comparison with Previous Methods}\label{subsec:comparemethod}
This subsection presents a thorough comparison between \backmc and previous methods. 

First, considering \setpush, it achieves the best-known worst-case complexity of $O\left(\log{n}\cdot \min\left(d_t, \sqrt{m}\right)\right)$ for this problem, with a hidden multiplicative factor of $1/\alpha^3$. In contrast, \backmc has a worst-case computational complexity of $O\left(\frac{1}{\dmin}\cdot \min\left(d_t, \sqrt{m}\right)\right)$, with a more favorable multiplicative dependency on $\alpha$ of $1/\alpha^2$. This complexity surpasses that of \setpush by a factor of $\Theta(\log{n}/\dmin)$. Additionally, the improvement in the $\alpha$ factor can be significant in practice when $\alpha$ is small (e.g., $\alpha=0.01$).
Furthermore, in Section~\ref{sec:MCanalysis}, we will demonstrate that the complexity bound of our \backmc is optimal, matching the lower bound of $\Omega\left(\frac{1}{\dmin} \cdot \min\left(d_t, \sqrt{m}\right)\right)$ for this problem. 
It is also worth noting that \backmc offers a much simpler algorithm and analysis compared to \setpush~\cite{setpush2023VLDB}. \setpush devises a complex push strategy to estimate $\vpi(t,s)$ for all $s\in V$, and then compute the approximation of $\vpi(t)$ using the equation $\vpi(t)=\frac{1}{n}\cdot \sum_{s\in V} \frac{d_t}{d_s}\cdot \vpi(t,s)$, as inferred from Equation~\eqref{eqn:PageRank_PPR} and~\eqref{eqn:birectional_ppr}. However, our analysis shows that this is unnecessary. To estimate $\vpi(t,s)$, generating $\alpha$-discounted random walks from node $t$ suffices to achieve optimal complexity. 

For other previous methods, we now analyze the reasons why these methods cannot achieve the optimal complexity results. First of all, the MC method~\cite{fogaras2005MC, avrachenkov2007monte, borgs2012sublinear, borgs2014multiscale} estimates $\vpi(t)$ directly, resulting in a computational complexity lower bound of $\Omega(n)$ in an $n$-node graph. Specifically, in an $n$-node graph $G$, all but $o(n)$ nodes have PageRank scores $O(1/n)$. Therefore, the MC method needs to generate $\Omega(n)$ $\alpha$-discounted random walks, each initiated from a uniformly random source node, in order to hit a target node $t$ with $\vpi(t)=O(1/n)$ at least once. For the push-based methods like \bpush~\cite{andersen2007contribution} and \bpushpeter~\cite{lofgren2013personalized}, they can perform $\Omega(n)$ elementary query operations during a push operation at a node $u$ with $d_u=O(n)$. The \unbippr method~\cite{lofgren2015bidirectional_undirected} combines the MC method and push operations, thus inheriting their drawbacks. The \rbs, \sublinear, and \sublinearplus methods are three exceptions, each of which devises the original push operation. However, \rbs~\cite{wang2020RBS}, \sublinear~\cite{bressan2018sublinear}, and \sublinearplus~\cite{bressan2023sublinear} are tailored for general directed graphs and do not leverage the symmetry of PPR scores in undirected graphs. This limitation prevents them from achieving optimality in undirected graphs due to the differences in local PageRank estimation between undirected and directed graphs. 


\section{Lower Bounds}
\label{sec:MCanalysis}

\begin{figure}[t]
\centering
\begin{tabular}{c}
\includegraphics[width=85mm,trim=0 0 0 0,clip]{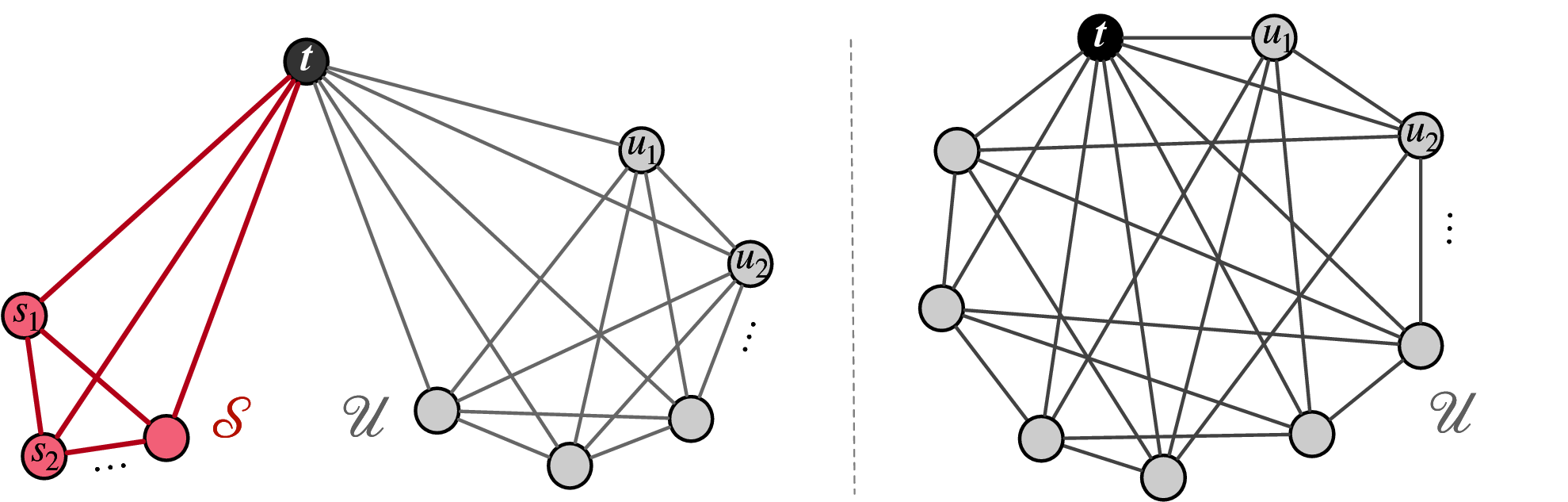}
\end{tabular}
\vspace{-2mm}
\caption{Hard instances of the lower bound proof. }
\label{fig:lowerbound}
\end{figure}

This section establishes the computational complexity lower bound for the local approximation of $\vpi(t)$. Specifically, we have Theorem~\ref{thm:lowerbound} as shown below. 
\begin{theorem}\label{thm:lowerbound}
Choose any integer $\vap\ge 2$ and any functions $m(n) \in \Omega(n)$, $\dmin(n)\in \Omega(1)\cap O(n)$. Consider any (randomized) algorithm $\calH(t)$ that estimates $\vpi(t)$ within a multiplicative factor $O(1)$ with probability at least $1/\vap$, where $\calH$ can only access the unseen nodes and edges in the underlying graph via a graph oracle. Then, for every sufficiently large $n$, there exists an undirected graph $G$ such that: 
\begin{enumerate}
\item The graph $G$ contains $\Theta(n)$ nodes and $\Theta(m)$ edges, and its minimum degree is $\Theta(\dmin)$; 
\item The graph $G$ contains a node $t$ with degree $d_t$, such that $\calH(t)$ requires $\Omega\left(\frac{1}{\dmin}\cdot \min\left(d_t, m^{1/2}\right)\right)$ elementary query operations in expectation in the arc-centric graph-access model. 
\end{enumerate}
\end{theorem}

Let us outline the proof first for ease of understanding. We establish Theorem~\ref{thm:lowerbound} by constructing $(\vap+1)$ graphs as hard instances for estimating $\vpi(t)$. These graphs are carefully designed so that any algorithm requires $\Omega\left(\frac{1}{\dmin}\cdot \min\left(d_t, m^{1/2}\right)\right)$ elementary query operations in the arc-centric graph access model to distinguish between them. However, distinguishing these graphs is essential for any algorithm to produce an approximation of $\vpi(t)$ within a multiplicative factor $O(1)$ with a probability of at least $1/p$. Hence, we establish a query complexity lower bound of $\Omega\left(\frac{1}{\dmin}\cdot \min\left(d_t, m^{1/2}\right)\right)$ for estimating $\vpi(t)$. As detailed in Section~\ref{subsec:graphmodel}, the query complexity of an algorithm is a lower bound for its computational complexity, thus establishing Theorem~\ref{thm:lowerbound}. 


We present the formal proof of Theorem~\ref{thm:lowerbound} in the following. 

\begin{proof}
We construct $(\vap+1)$ graphs $G^{(0)}, G^{(1)}, \ldots, G^{(\vap)}$ as hard instances for estimating $\vpi(t)$. In graph $G^{(i)}$, the target node $t$ has $\Theta\left(d_t\right)$ neighbors, which can be divided into $(i+1)$ groups. Each of the first $i$ groups, denoted by $\mathcal{S}^{(j)}$ for $j\in [0, i)$, has $\Theta(\dmin)$ neighbor nodes, while the last group, denoted by $\mathcal{U}$, contains $\Theta(m^{1/2})$ neighbors. In the first $i$ groups, each neighbor node has a degree of $\Theta(\dmin)$, whereas in the $(i+1)$-th group, each neighbor node has a degree of $\Theta(m^{1/2})$. Additionally, we add isolated nodes to these graphs, ensuring that the number of nodes in each graph is $\Theta(n)$. Figure~\ref{fig:lowerbound} illustrates graphs $G^{(1)}$ and $G^{(0)}$.

We denote the PageRank score of $t$ in graph $G^{(i)}$ as $\vpi_i(t)$. We will demonstrate that for each $i \in [1,\vap]$, 
$$\vpi_i(t)=(1+\Omega(1))\cdot \vpi_{i-1}(t).$$
Specifically, in graph $G^{(i)}$, each node $s$ belonging to group $\mathcal{S}^{(j)}$ for $j\in [0, i)$ has a Personalized PageRank (PPR) score $\vpi(s,t)=\Theta(1/\dmin)$. Furthermore, each node $u\in \mathcal{U}$ in graph $G^{(i)}$ has a PPR score $\vpi(u,t)=\Theta(1/m^{1/2})$. Additionally, the PPR score of node $t$ to itself is $\vpi(t,t)=\Theta(1/n)$. Thus, according to Equation~\eqref{eqn:PageRank_PPR}, the PageRank score of node $t$ in graph $G^{(i)}$ is 
\begin{align*}
\vpi_i(t)=\Theta\left(\frac{1}{n}\left(1+i \cdot \dmin \cdot \frac{1}{\dmin} +m^{1/2} \cdot \frac{1}{m^{1/2}}\right)\right)=\Theta\left(\frac{1}{n}\right).	
\end{align*}
The difference $\vpi_i(t)-\vpi_{i-1}(t)=\Theta\left(\frac{1}{n}\cdot \dmin \cdot \frac{1}{\dmin}\right)=\Theta\left(\frac{1}{n}\right)$, which corroborates the assertion that 
$\vpi_i(t)=(1+\Omega(1))\cdot \vpi_{i-1}(t)$
for each $i \in [1,\vap]$. 

Furthermore, let $\G^{(i)}$ represent the set of all $n!$ graphs that are isomorphic to $G^{(i)}$ obtained through permutation of node labels. 
Consider an undirected graph $G$, chosen uniformly at random from $\bigcup_{i=1}^{\vap} \G^{(i)}$. Any algorithm $\calH(t)$ is required to identify the specific set $\G^{(i)}$ from which $G$ originates. Failing this, the probability that $\calH(t)$ can estimate $\vpi(t)$ within a multiplicative factor of $O(1)$ diminishes to at most $1/(\vap+1)$, a result deemed unacceptable.

Finally, in the arc-centric graph access model, we contend that any algorithm $\calH(t)$ must execute $\Omega\left(\frac{1}{\dmin}\cdot \min\left(d_t, m^{1/2}\right)\right)$ queries in expectation to distinguish between graphs $G^{(i-1)}$ and $G^{(i)}$ for any $i \in [1,\vap]$. By our design, $\calH(t)$ must detect at least one node in group $\mathcal{S}^{(i-1)}$ to discern the difference between graphs $G^{(i-1)}$ and $G^{(i)}$, given that other sections of the two graphs remain indistinguishable. It is important to note that $\calH(t)$ can explore unseen nodes in the underlying graph only through interactions with the graph oracle, utilizing the three elementary query operations: $\deg(\cdot)$, $\neighbor(\cdot,\cdot)$, and $\jump()$. 
When employing the $\jump()$ query, $\calH(t)$ is expected to perform $\Omega(n/\dmin)$ $\jump$ queries to detect a node in group $\mathcal{S}^{(i-1)}$ with probability at least $1/\vap$. However, this complexity, $\Omega(n/\dmin)$, exceeds the stated lower bound asymptotically. Consequently, the viable strategy for $\calH(t)$ is only local access through $\neighbor(t,\cdot)$ and $\deg(\cdot)$ queries. This approach requires $\calH(t)$ to perform $\Omega\left(\frac{m^{1/2}}{\dmin}\right)$ queries on average to detect at least one node in $\mathcal{S}^{(i-1)}$. Notably, $\Theta\left(\frac{m^{1/2}}{\dmin}\right) = \Theta\left(\frac{d_t}{\dmin}\right)$, given that $d_t=\Theta(\dmin+m^{1/2}) = \Theta(m^{1/2})$ holds in these hard instances. This substantiates the proposed lower bound of $\Omega\left(\frac{1}{\dmin} \cdot \min\left(d_t, m^{1/2}\right)\right)$, thus concluding the proof. 
\end{proof}



\section{Experiments} \label{sec:exp}

We evaluate our \backmc algorithm against other baseline methods on large-scale real-world and synthetic graphs. 

\vspace{-1mm}
\begin{table}[t]
\vspace{-2mm}
\centering
\caption{Datasets}
\vspace{-2mm}
\begin{small}
\begin{tabular}
{llrrr} 
\toprule
{{\bf Dataset}}& {{\bf Graph Type}}& {{\bf $\boldsymbol{n}$}} & {{\bf $\boldsymbol{m}$}} & {{\bf  $\boldsymbol{\dmin}$}} \\ \midrule
{Youtube(YT)} & {Real-World} & {1,138,499} & {5,980,886} & {1}\\
{LiveJournal (LJ)} & {Real-World} & {4,847,571} & {85,702,474} & {1}\\
{Twitter (TW)} & {Real-World} & {41,652,230} & {2,405,026,092} & {1}  \\ 
{Friendster (FR)} & {Real-World} &{68,349,466} & {3,623,698,684} & {1} \\ \midrule
{ER10} & {Synthetic} & {100,000} & {1,001,008} & {1}  \\ 
{ER100} & {Synthetic} & {100,000} & {9,993,692} & {59}  \\ 
{ER1000} & {Synthetic} & {100,000} & {100,001,498} & {875}  \\ 
{ER10000} & {Synthetic} & {100,000} & {1,000,052,806} & {9582}  \\ 
\bottomrule
\end{tabular}
\end{small}
\label{tbl:datasets}
\vspace{-2mm}
\end{table}

\header{\bf Environment. } We conduct all the experiments on a Linux server with an Intel (R) Xeon(R) Gold 6126@2.60GHz CPU and 500GB memory. All the methods are implemented in C++ and compiled in g++ with the O3 optimization turned on. 

\header{\bf Methods and Parameters. } We compare our \backmc against six baseline methods, including \mc~\cite{fogaras2005MC}, \bpushpeter~\cite{lofgren2013personalized}, \rbs~\cite{wang2020RBS}, \setpush~\cite{setpush2023VLDB}, \unbippr~\footnote{It's worth noting that \unbippr~\cite{lofgren2015bidirectional_undirected} is initially implemented as a combination of forward push and MC, which however is impractical under the graph access model due to the $O(n)$ time required to identify all nodes for pushing in the initial state. Instead, we implement \unbippr by combining backward push and MC, ensuring an $O(n^{1/2}d_t^{1/2})$ complexity under the graph access model.}~\cite{lofgren2015bidirectional_undirected}, and \sublinearplus~\cite{bressan2023sublinear}. Additionally, we compute the ground truths by iteratively updating the PageRank vector $\vpi$ using Equation~\eqref{eqn:def_pagerank}. Initially, we set $\vpi=\bm{1}/n$ and repeat the iteration $\lceil \log_{1-\alpha}{\frac{0.0001 \cdot \alpha}{n}}\rceil$ times. Then for any given target node $t$, we utilize the derived $\vpi(t)$ as the ground truth of the PageRank score of $t$. Throughout our experiments, we set the failure probability $\pf$ to $0.1$ and vary the relative error $\rela$ within the range $(0.01, 0.5]$ to analyze the tradeoff between query time and actual {Relative Error} (as defined below) for each method. Moreover, unless otherwise specified, we set the teleport probability $\alpha$ to $0.2$.

\header{\bf Metrics. } We consider actual {\em Relative Error}, which is defined as
\begin{align*}
\textit{Relative Error}=\frac{1}{\vpi(t)}\cdot \left|\vpi(t)-\epi(t)\right|. 
\end{align*}
On each dataset, we sample $10$ nodes as the target node $t$, and execute each method once for each target node. Subsequently, we compute the average actual \textit{Relative Error} of each method for estimating $\vpi(t)$ across all query nodes under each setting of $\rela$.

\begin{figure*}[t]
\begin{minipage}[t]{1\textwidth}
\centering
\begin{tabular}{cccc}
\hspace{-4mm} \includegraphics[width=45mm]{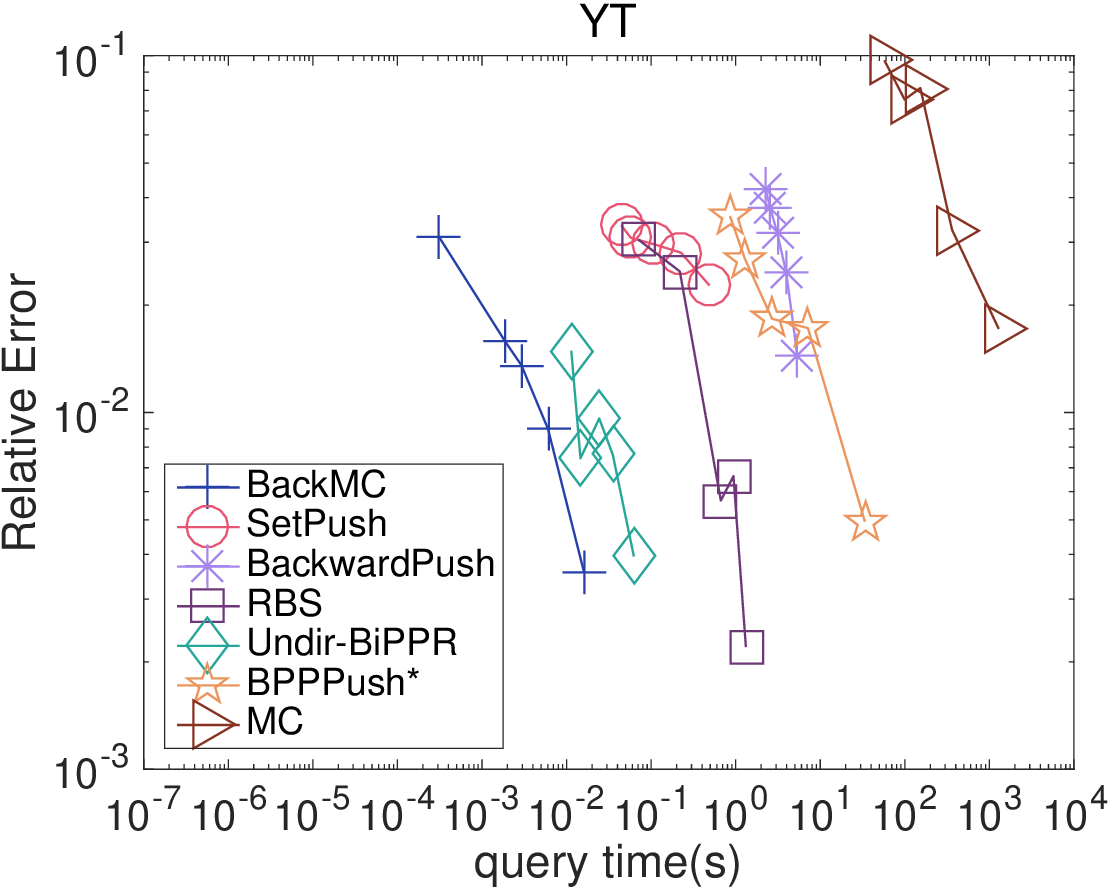} &
\hspace{-3mm} \includegraphics[width=45mm]{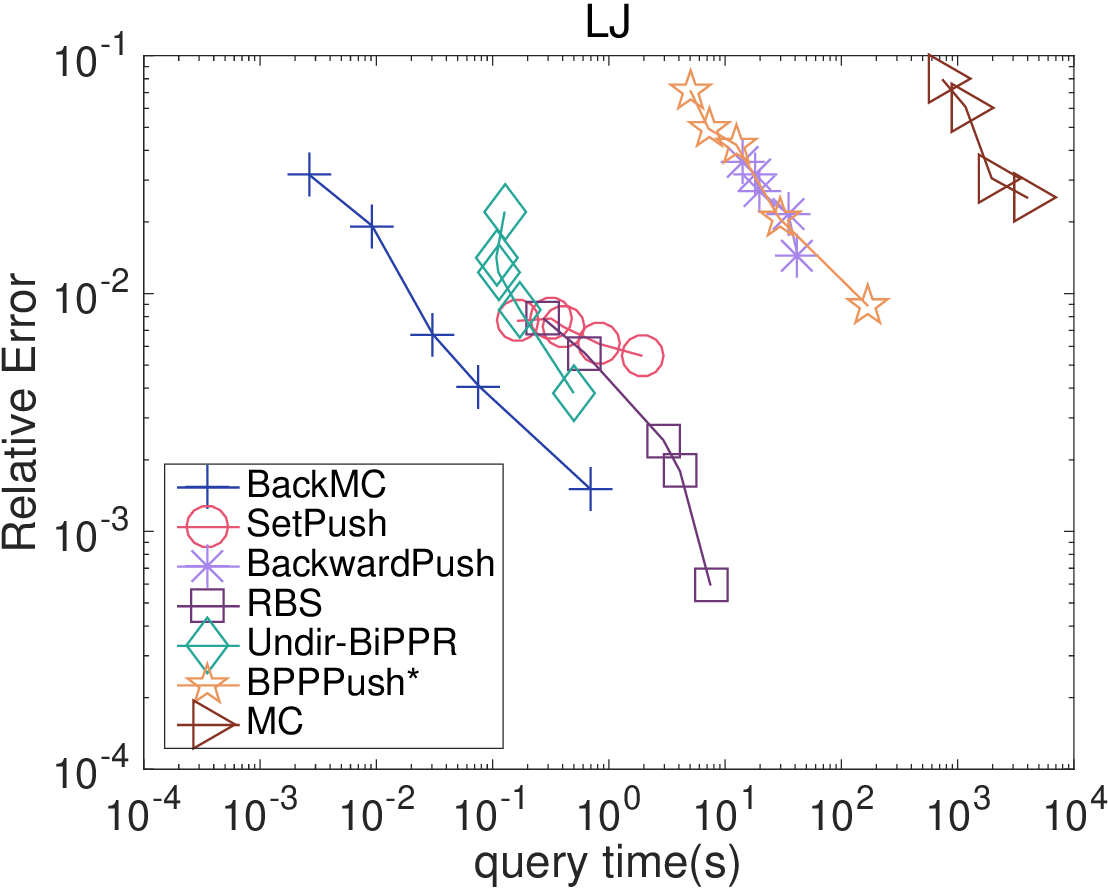} &
\hspace{-3mm} \includegraphics[width=45mm]{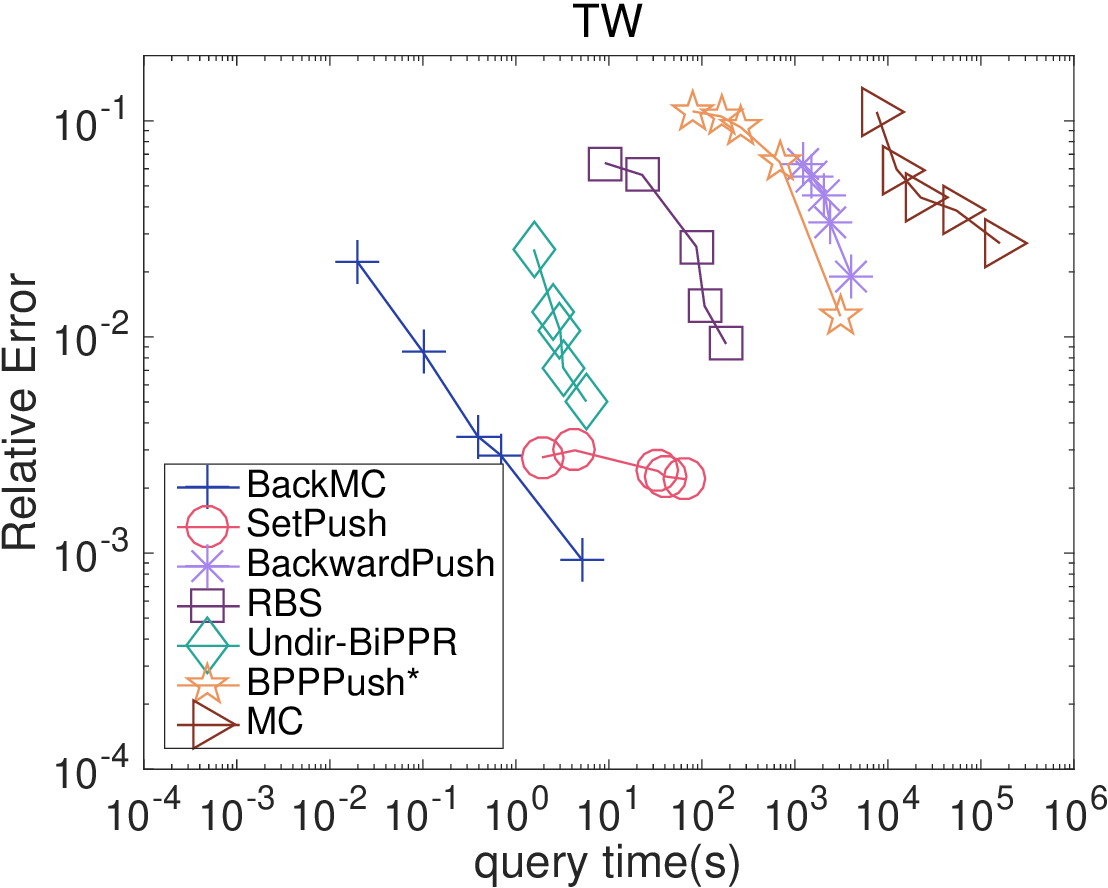} &
\hspace{-3mm} \includegraphics[width=45mm]{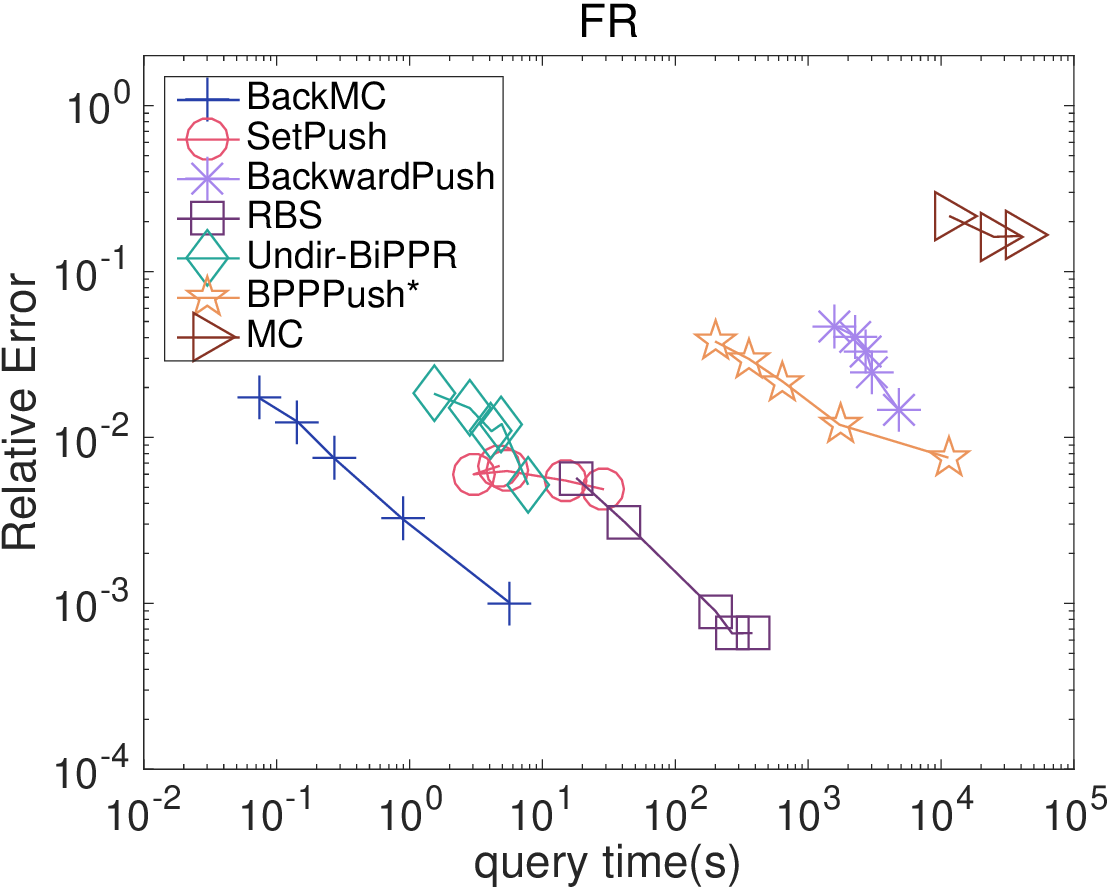} \\
\end{tabular}
\vspace{-3mm}
\caption{actual Relative Error v.s. query time (seconds), the target node $\boldsymbol{t}$ sampled uniformly, $\alpha=0.2$}
\label{fig:query_uniform}
\vspace{+2mm}
\end{minipage}

\begin{minipage}[t]{1\textwidth}
\centering
\begin{tabular}{cccc}
\hspace{-4mm} \includegraphics[width=45mm]{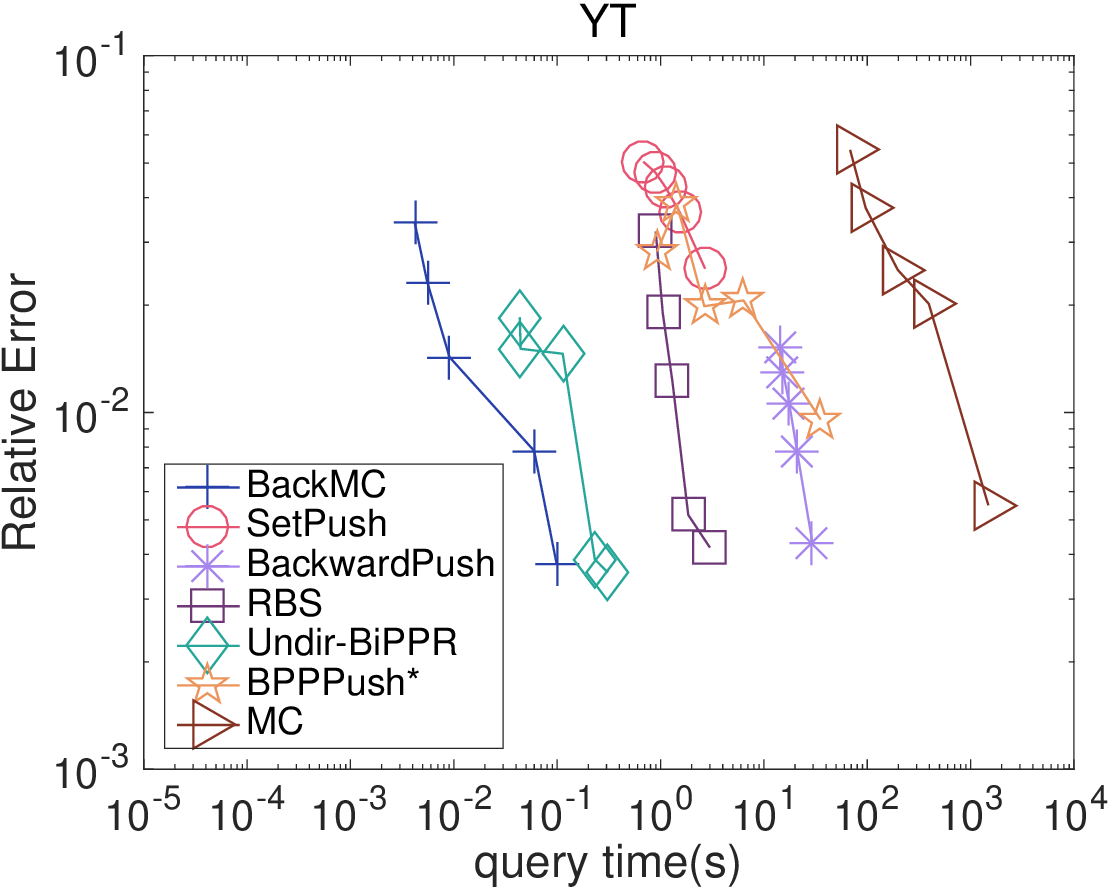} &
\hspace{-3mm} \includegraphics[width=45mm]{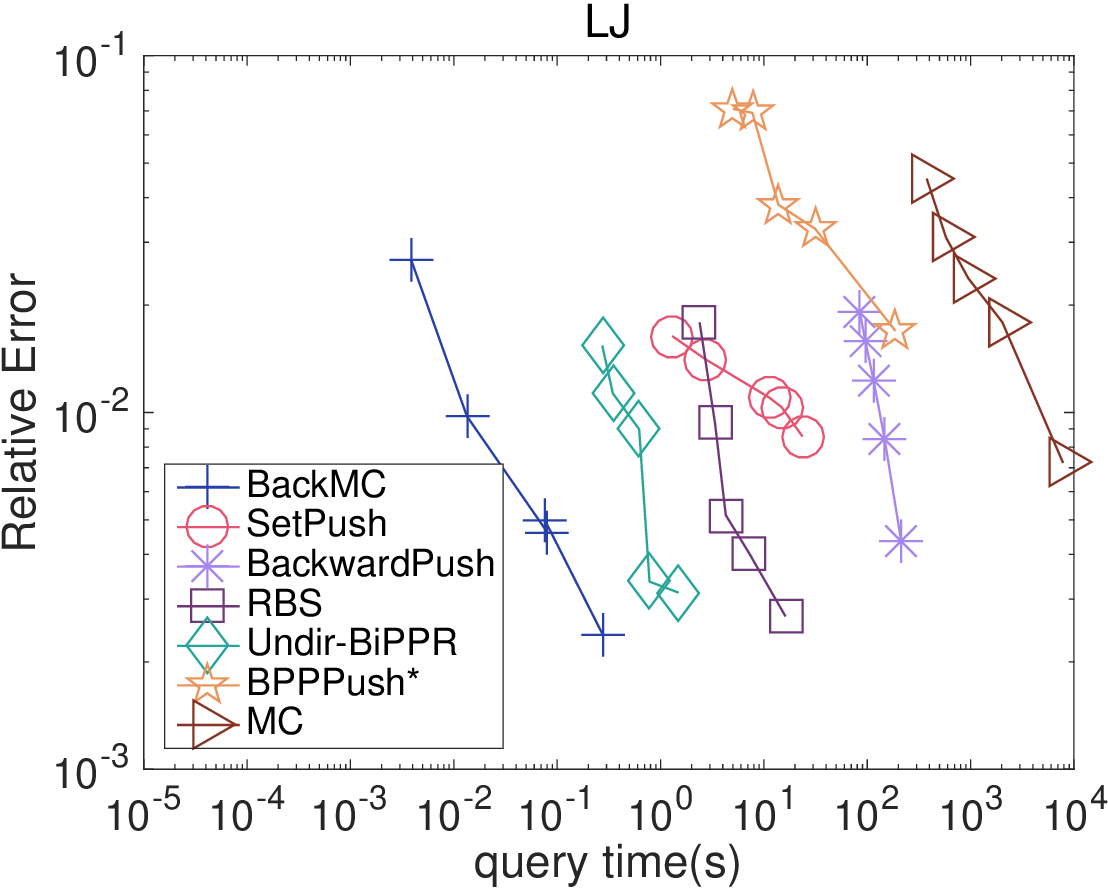} &
\hspace{-3mm} \includegraphics[width=45mm]{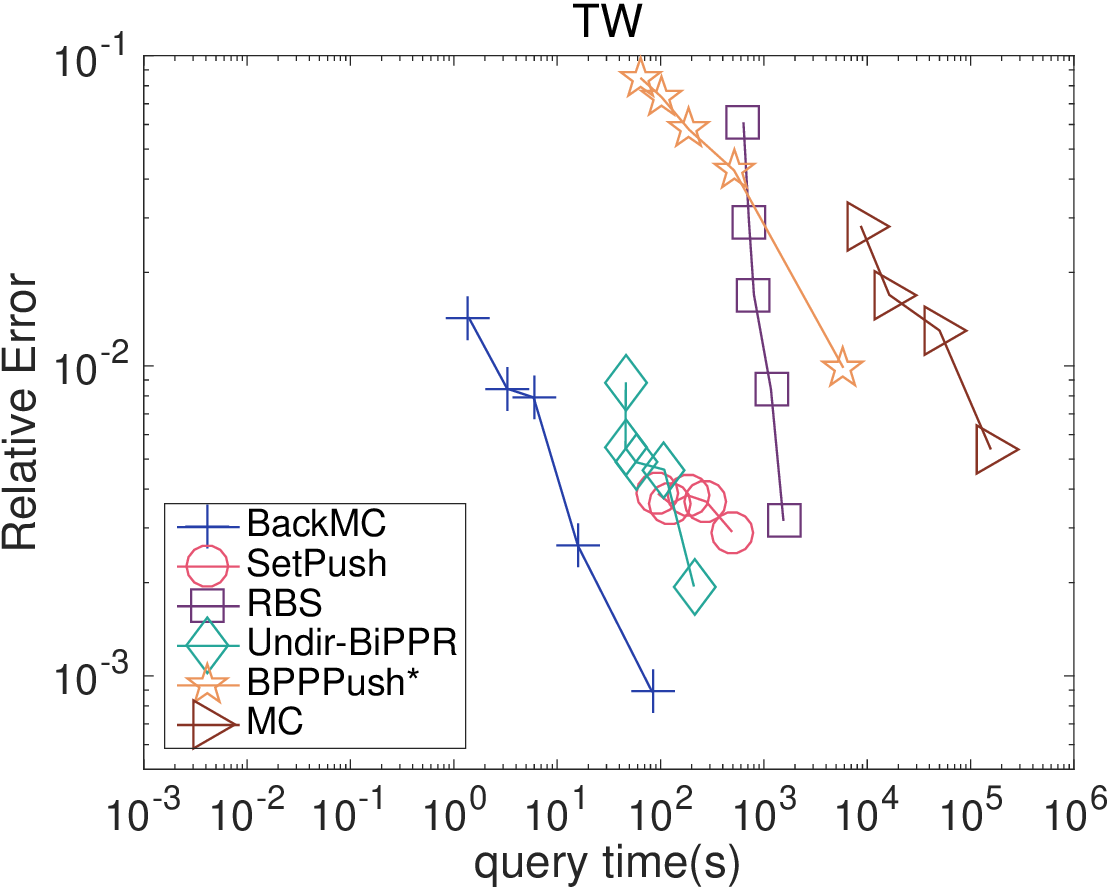} &
\hspace{-3mm} \includegraphics[width=45mm]{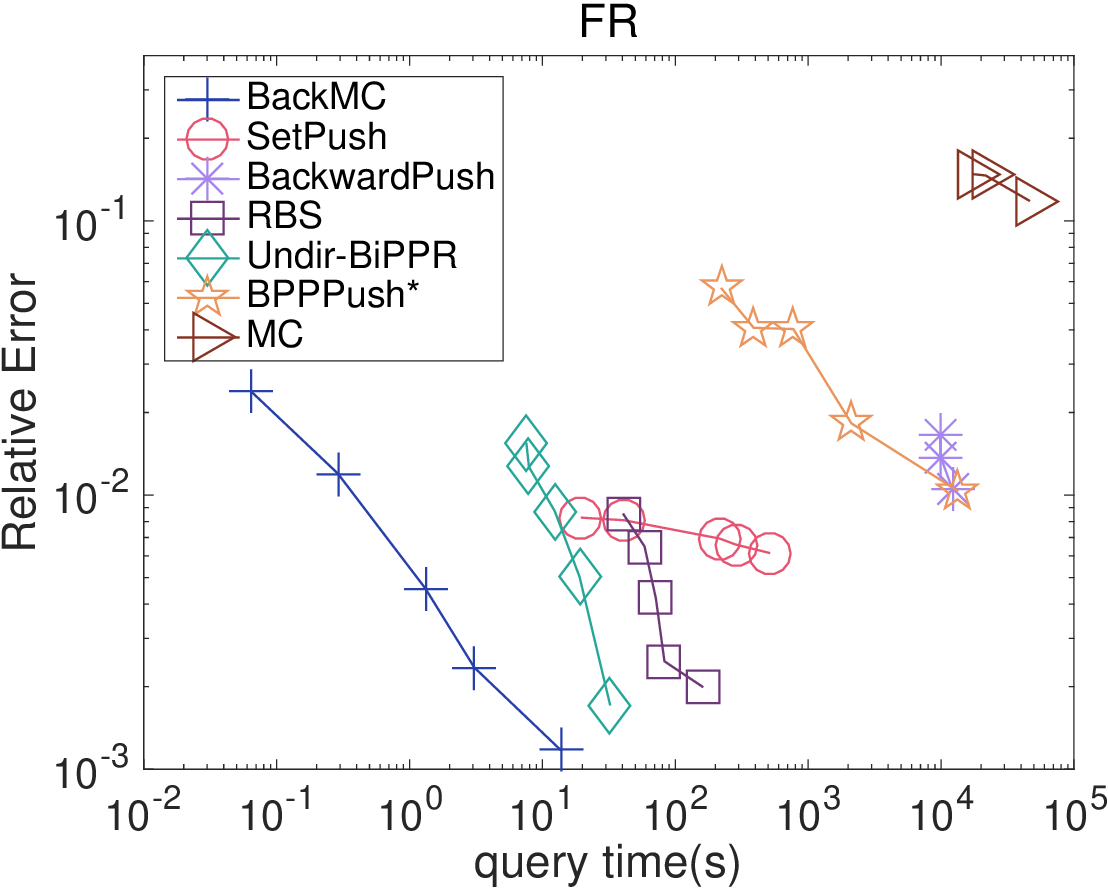} \\
\end{tabular}
\vspace{-3mm}
\caption{actual Relative Error v.s. query time (seconds), the target node $\boldsymbol{t}$ sampled from the degree distribution, $\alpha=0.2$}
\label{fig:query_degree}
\vspace{+2mm}
\end{minipage}

\begin{minipage}[t]{1\textwidth}
\centering
\begin{tabular}{cccc}
\hspace{-4mm} \includegraphics[width=45mm]{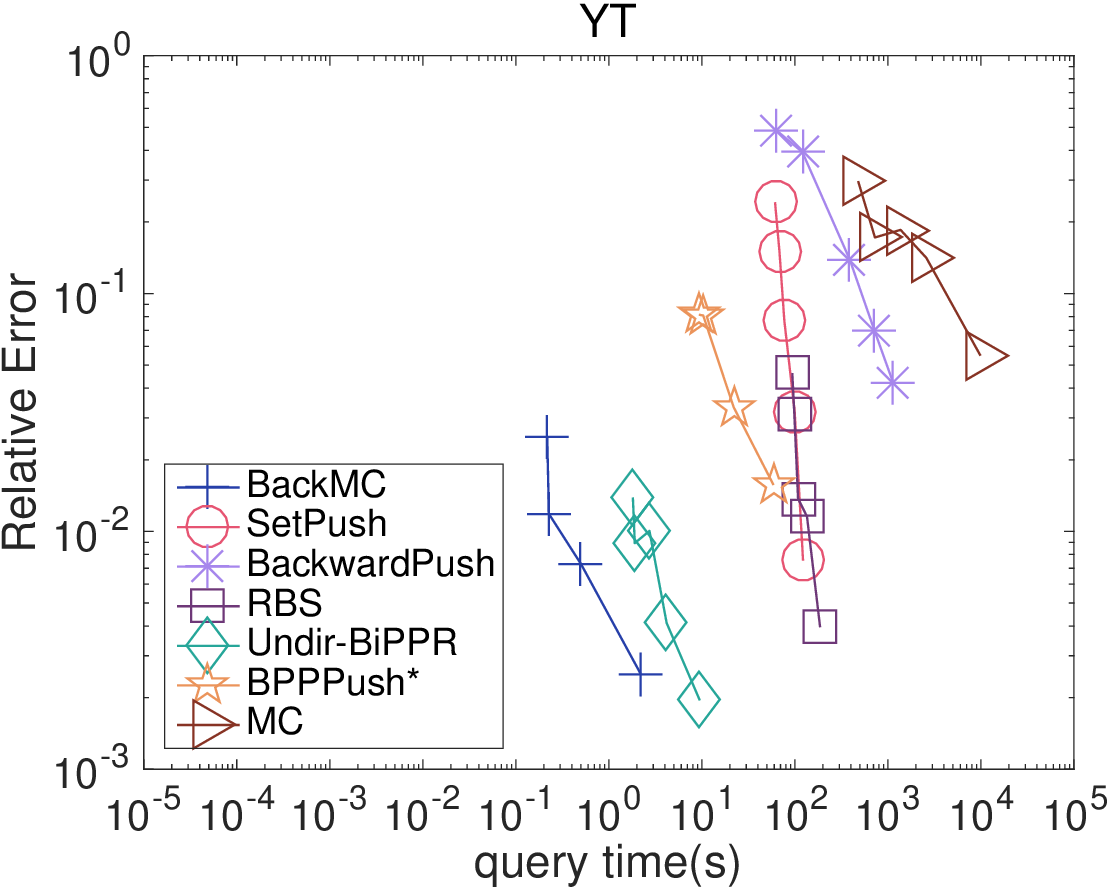} &
\hspace{-3mm} \includegraphics[width=45mm]{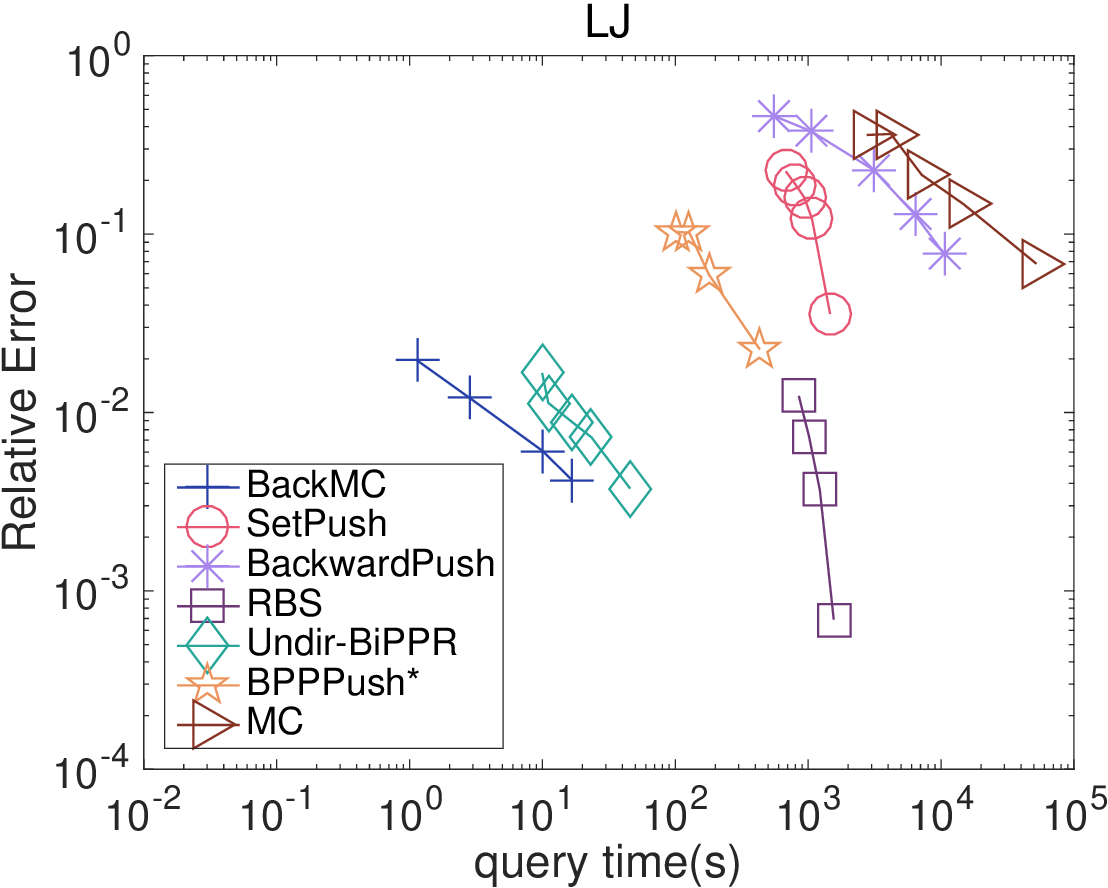} &
\hspace{-3mm} \includegraphics[width=45mm]{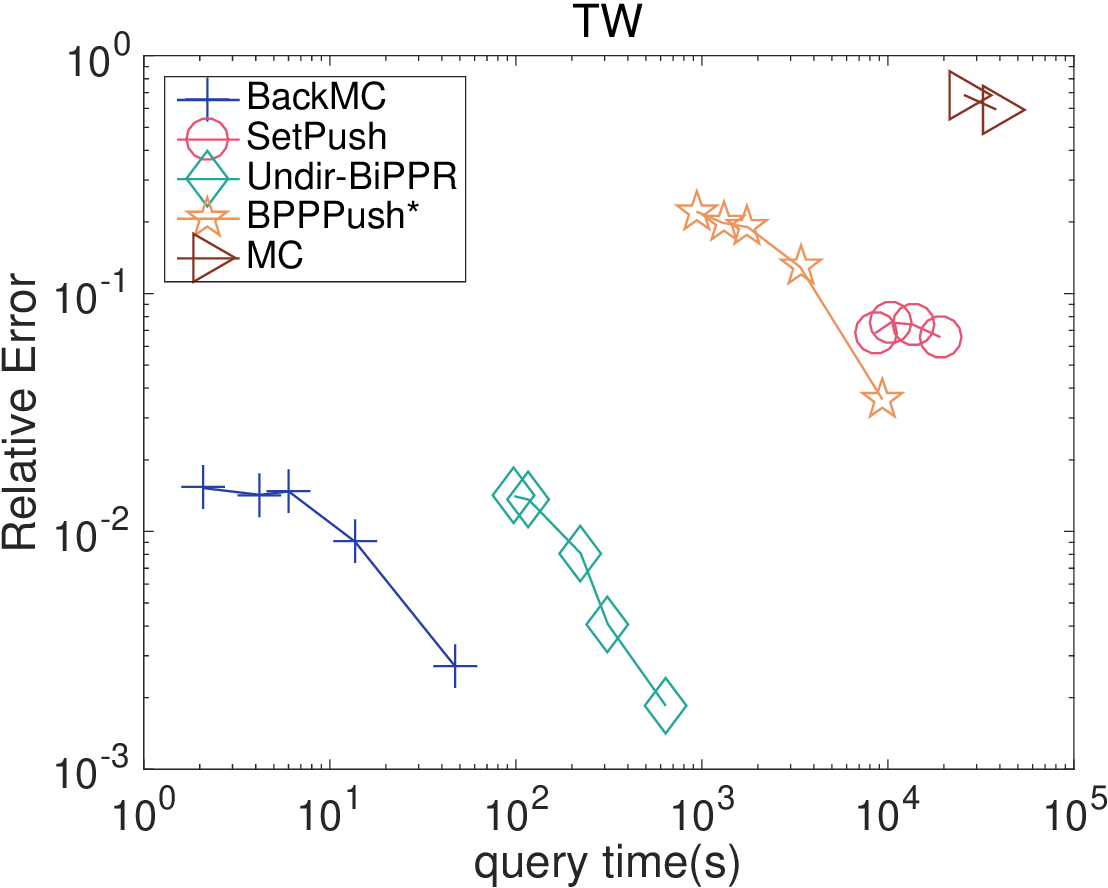} &
\hspace{-3mm} \includegraphics[width=45mm]{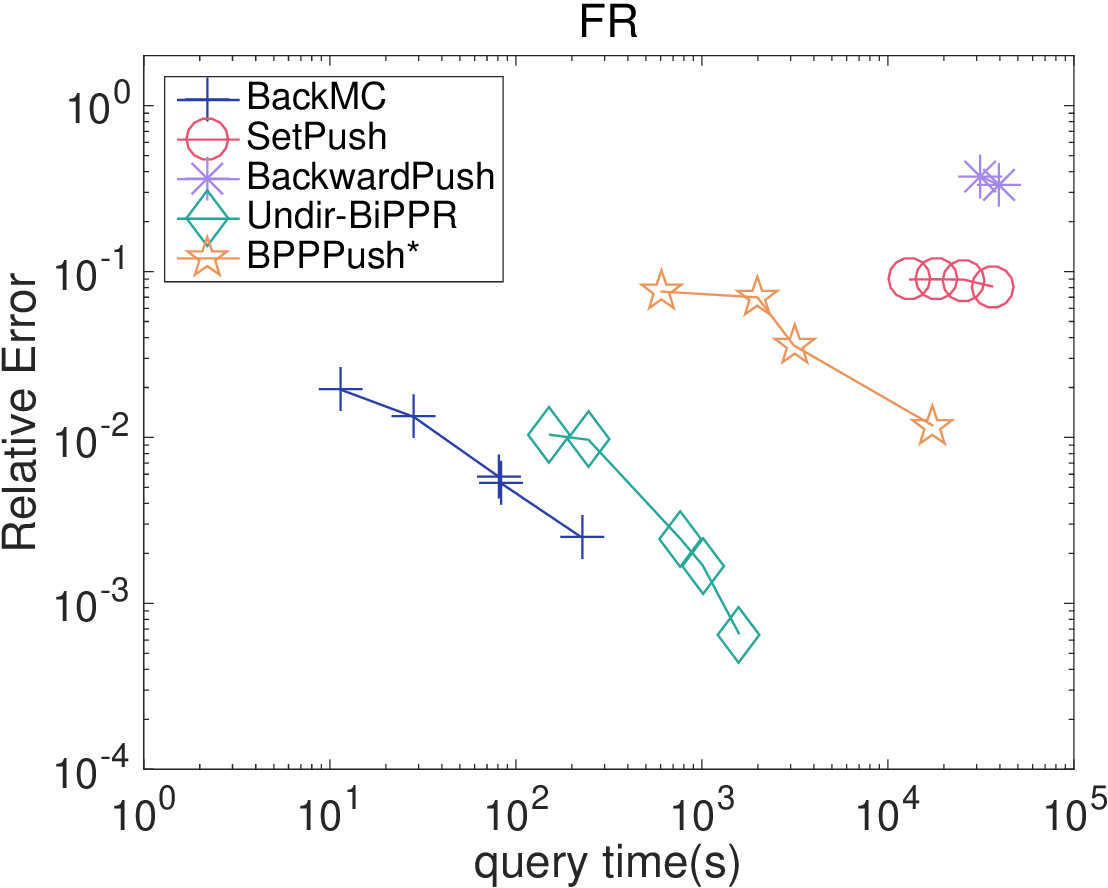} \\
\end{tabular}
\vspace{-3mm}
\caption{actual Relative Error v.s. query time (seconds), the target node $\boldsymbol{t}$ sampled uniformly, $\alpha=0.01$}
\label{fig:query_uniform_01}
\end{minipage}
\vspace{-1mm}
\end{figure*}

\subsection{Experiments on Real-World Graphs}
We first evaluate our \backmc against other baseline methods on large-scale real-world graphs.

\header{\bf Datasets. } We utilize four real-world datasets in our experiments: YouTube (YT), LiveJournal (LJ), Twitter (TW), and Friendster (FR). These datasets originate from social networks and are publicly available~\footnote{\url{http://law.di.unimi.it/datasets.php}}~\footnote{\url{http://snap.stanford.edu/data}}. In these graphs, nodes represent users on the respective websites, and edges denote friendships between users. Table~\ref{tbl:datasets} presents the statistics of the four datasets. 

\header{\bf Target Nodes. } For each dataset, we sample two subsets from the node set $V$ of $G$, each containing $10$ nodes designated as target nodes. In the first subset, nodes are randomly selected from $V$ with a uniform distribution. In contrast, nodes in the second subset are chosen from $V$ based on the degree distribution. Specifically, for any node $u \in V$, the likelihood of $u$ being sampled in the second subset increases with its degree $d_u$. This sampling strategy allows us to evaluate the performance of all methods in estimating $\vpi(t)$ for nodes with high degrees.

\header{\bf Results. } 
Figure~\ref{fig:query_uniform} shows the tradeoffs between query time and actual Relative Error for each method, with the target node $t$ uniformly selected from $V$. We note that \backmc consistently achieves the shortest query time compared to the other baseline methods at the same actual Relative Error level. Specifically, \backmc outperforms \unbippr, \setpush, and \rbs by a factor of $10\times$, and surpasses \sublinearplus, \mc, and \bpushpeter by $100\times$ to $1000\times$, in general. Particularly noteworthy is the significant superiority of \backmc over \setpush, which previously established the best complexity bound for estimating $\vpi(t)$. This empirical observation aligns with our theoretical arguments, not only showcasing the $\Theta\left(\frac{\log{n}}{\dmin}\right)$ theoretical improvement of \backmc over \setpush but also highlighting the clean algorithm structure and ease of implementation. Additionally, as mentioned in Section~\ref{subsec:comparemethod}, the presence of a multiplicative factor $1/\alpha^3$ in the complexity bound of \setpush hampers its empirical performance efficiency.

Figure~\ref{fig:query_degree} illustrates the tradeoffs between query time and actual Relative Error for each method, with the target node $t$ sampled from the degree distribution. Analogously, \backmc consistently outperforms all baseline methods on all datasets in terms of both efficiency and accuracy. Notably, we omit the \bpushpeter method from the Twitter dataset analysis due to its query time exceeding 12 hours. This observation aligns with our analysis that considerable time is required by \bpushpeter to perform push operations on nodes with high degrees, given that the Twitter dataset is dense with a large average degree, as shown in Table~\ref{tbl:datasets}.

In Figures~\ref{fig:query_uniform} and~\ref{fig:query_degree}, we maintain a fixed teleport probability $\alpha$ of 0.2. In Figure~\ref{fig:query_uniform_01}, we present the tradeoffs between query time and actual Relative Error with $\alpha=0.01$. Target nodes $t$ are uniformly sampled from $V$. Comparing the results from Figure~\ref{fig:query_uniform_01} and Figure~\ref{fig:query_uniform}, we observe a more pronounced superiority of \backmc over \setpush. This aligns well with our analysis that the \setpush method has a multiplicative factor of $1/\alpha^3$ in its complexity bound, which will become significant when $\alpha$ is small. In contrast, the computational complexity of our \backmc has a multiplicative $1/\alpha^2$ factor, which is more favorable than that of \setpush. Additionally, a small $\alpha$ also makes the query time of \rbs unaffordable on large-scale Twitter and Friendster datasets. We omit them from Figure~\ref{fig:query_uniform_01}. We also exclude \bpushpeter on Twitter and \mc on Friendster from Figure~\ref{fig:query_uniform_01}, as their query time both exceed 12h.

\subsection{Experiments on Synthetic Graphs}
In this subsection, we evaluate \backmc and the competitors on synthetic Erdős--R\'{e}nyi (ER) graphs. 

\header{\bf Datasets. } We generate four ER graphs, each comprising a fixed number of nodes ($n=100,000$). The first graph, designated as ER10, features an edge-connection probability of $10/n$ between any two nodes. In other words, the probability that any given pair of nodes in the ER10 graph will be connected is consistently set to $10/n$. This connection probability is sequentially increased to $100/n$, $1000/n$, and $10000/n$ for the subsequent graphs, referred to as ER100, ER1000, and ER10000, respectively. As a direct outcome, the average degree of nodes in the ER10 graph is approximately $10$, while the corresponding average degrees in the ER100, ER1000, and ER10000 graphs are roughly $100$, $1000$, and $10000$, respectively. Notably, we present the minimum degree (i.e., $\dmin$) observed in these four ER graphs in Table~\ref{tbl:datasets}, which are around $1$, $10$, $10^2$, and $10^3$ for each graph, in ascending order of connection probability. These variations in ER graphs serve as a basis to demonstrate the superior performance of \backmc across a spectrum of graphs distinguished by differing values of $\dmin$. 

\begin{figure*}[t]
\centering
\begin{tabular}{cccc}
\hspace{-4mm} \includegraphics[width=45mm]{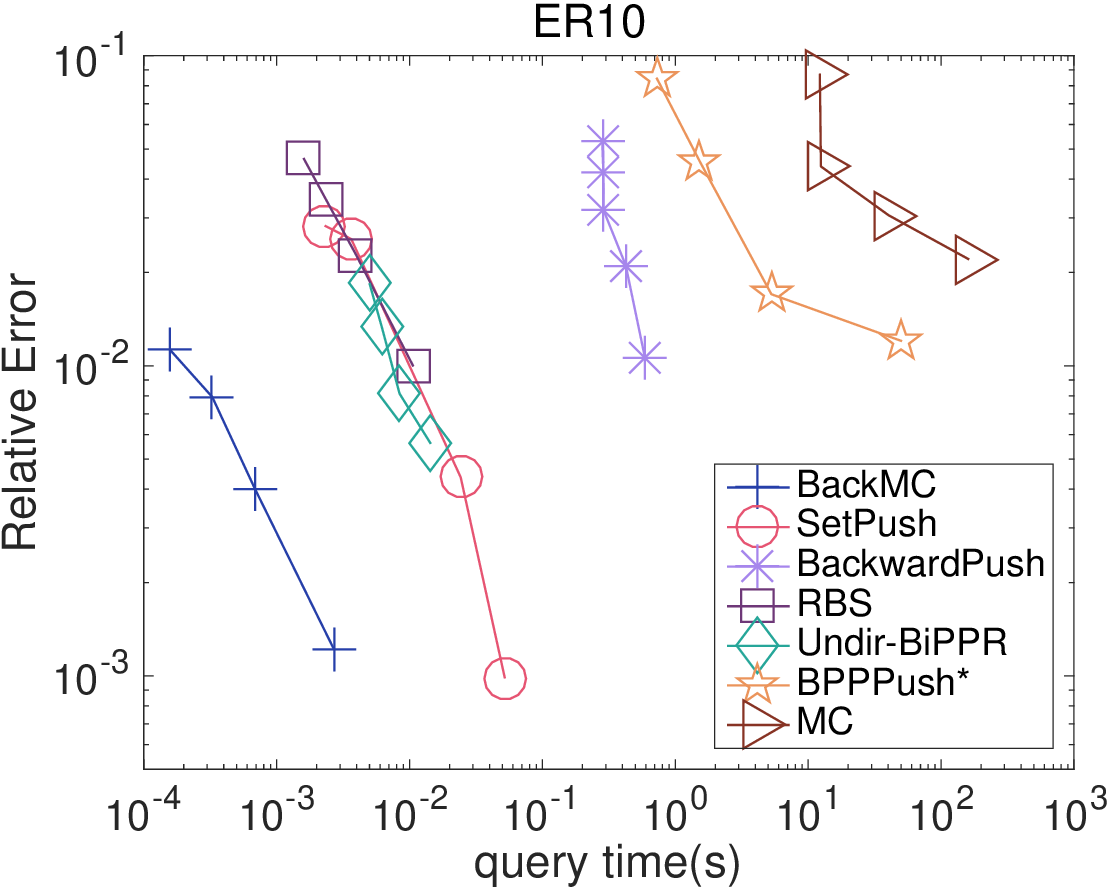} &
\hspace{-3mm} \includegraphics[width=45mm]{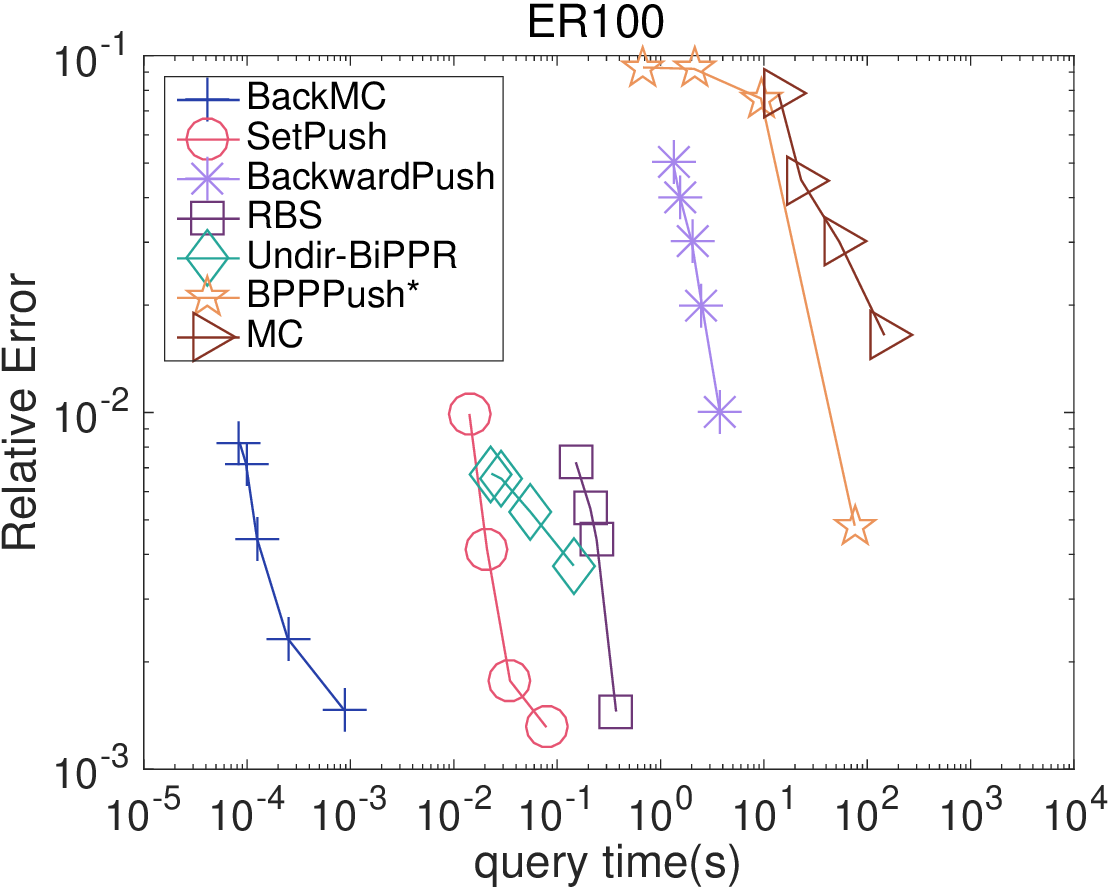} &
\hspace{-3mm} \includegraphics[width=45mm]{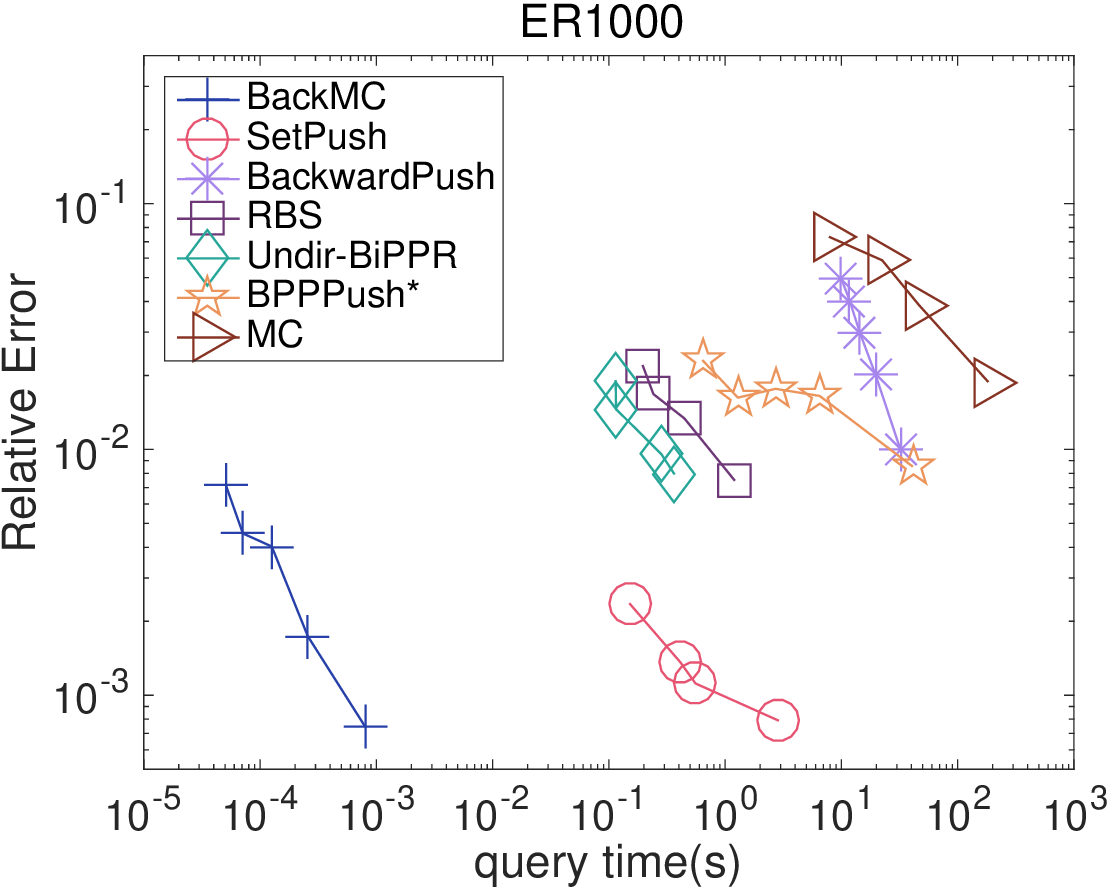} &
\hspace{-3mm} \includegraphics[width=45mm]{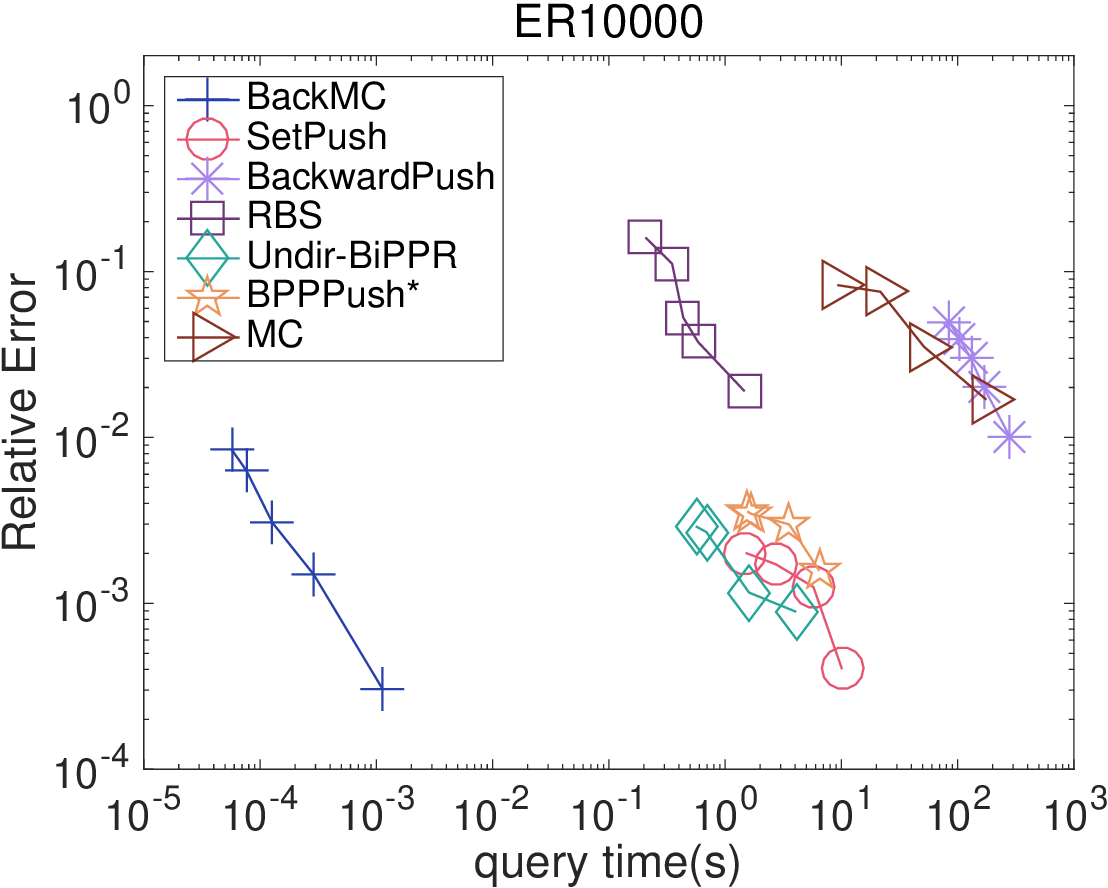} \\
\end{tabular}
\vspace{-3mm}
\caption{actual Relative Error v.s. query time (seconds) on synthetic graphs, the target node $\boldsymbol{t}$ sampled uniformly, $\alpha=0.2$}
\label{fig:query_ER}
\end{figure*}

\header{\bf Results. } In Figure~\ref{fig:query_ER}, we present the trade-off between actual relative error and query time across the four ER graphs. Our first observation is that \backmc consistently surpasses all baseline methods by orders of magnitude in query time for the same level of actual relative error. Notably, the margin of \backmc's superiority over its competitors increases as $\dmin$ rises. Specifically, on the ER10 graph, \backmc is tenfold faster than both \setpush, \unbippr and \rbs. This leading factor escalates from tenfold to a thousandfold. On the ER10000 graph, \backmc exceeds the performance of all competitors by at least three orders of magnitude in query time for comparable actual relative errors. Additionally, it is important to highlight that while the query time for competitor methods escalates with an increase in $\dmin$ (thus attributable to the growing number of edges), \backmc uniquely exhibits reduced query times as $\dmin$ increases. This phenomenon underscores \backmc's exceptional efficiency and corroborates our analysis, demonstrating a negative correlation between the computational complexity of \backmc and $\dmin$.


\vspace{-2mm}
\section{Conclusion} \label{sec:conclusion}
This paper introduces a simple and optimal algorithm, \backmc, for estimating a single node's PageRank score in undirected graphs. We have demonstrated the optimality of \backmc and assessed its performance on large-scale graphs. As for future directions, we aim to adapt \backmc to tackle single-pair PPR queries in undirected graphs. 
Given that the computation of PPR scores relates closely to that of PageRank scores, as shown in Equation~\eqref{eqn:PageRank_PPR}, we are motivated to investigate further in this area.
%


\begin{acks}
I would like to thank Professor Zhewei Wei for his unconditional support. I would also like to thank the anonymous reviewers for their insightful comments. 
\end{acks}

\balance
\bibliographystyle{plain}
\bibliography{paper}


\end{document}